\documentclass[lettersize,journal]{IEEEtran}

\ifCLASSOPTIONcompsoc
  \usepackage[nocompress]{cite}
\else
  \usepackage{cite}
\fi

\usepackage{markdown}
\usepackage{listings}
\usepackage{url}
\usepackage{graphicx}
\usepackage{subfig}
\usepackage{cite}
\usepackage{comment}
\usepackage{caption}
\usepackage{tabularx}
\usepackage{multirow}
\usepackage{multicol}
\usepackage{diagbox}
\usepackage{amsmath}
\usepackage{makecell}
\usepackage{booktabs}
\usepackage{spark}
\usepackage{sparklines}
\usepackage{color, colortbl}
\usepackage{soul}
\usepackage{arydshln}

\usepackage{algorithm}
\usepackage{algpseudocode}
\usepackage{amsfonts}
\usepackage{gensymb}

\usepackage{tcolorbox}
\newcommand{\revised}[1]{{\color{black} #1}}

\usepackage{enumitem}

\usepackage{xspace}

\usepackage[framemethod=TikZ]{mdframed}
\definecolor{mycolor}{RGB}{224,215,188}
\definecolor{bgcolor}{RGB}{249,245,233}

\newmdenv[
  backgroundcolor=bgcolor,
  topline=false,
  bottomline=false,
  rightline=false,
  leftline=false,
  skipabove=0.5em,
  skipbelow=0.5em,
  innerleftmargin=0.5em,
  innerrightmargin=0.5em,
  innertopmargin=0.5em,
  innerbottommargin=0.5em,
  singleextra={
    \draw[mycolor, line width=0.5pt] ([yshift=0pt]O) -- ([yshift=0pt]O-|P);
    \draw[mycolor, line width=0.5pt] ([yshift=0pt]P-|O) -- ([yshift=0pt]P);
    \draw[mycolor, line width=0.5pt] ([yshift=1.5pt]O) -- ([yshift=1.5pt]O-|P);
    \draw[mycolor, line width=0.5pt] ([yshift=-1.5pt]P-|O) -- ([yshift=-1.5pt]P);
  },
  secondextra={
    \draw[mycolor, line width=0.5pt] ([yshift=0pt]O) -- ([yshift=0pt]O-|P);
    \draw[mycolor, line width=0.5pt] ([yshift=1.5pt]O) -- ([yshift=1.5pt]O-|P);
  },
  firstextra={
    \draw[mycolor, line width=0.5pt] ([yshift=0pt]P-|O) -- ([yshift=0pt]P);
    \draw[mycolor, line width=0.5pt] ([yshift=-1.5pt]P-|O) -- ([yshift=-1.5pt]P);
  }
]{coloredquotation}

\definecolor{output}{RGB}{222,234,240}
\definecolor{mycoloroutput}{RGB}{177,211,229}
\newmdenv[
  backgroundcolor=output,
  topline=false,
  bottomline=false,
  rightline=false,
  leftline=false,
  skipabove=0.5em,
  skipbelow=0.5em,
  innerleftmargin=0.5em,
  innerrightmargin=0.5em,
  innertopmargin=0.5em,
  innerbottommargin=0.5em,
    singleextra={
    \draw[mycoloroutput, line width=0.5pt] ([yshift=0pt]O) -- ([yshift=0pt]O-|P);
    \draw[mycoloroutput, line width=0.5pt] ([yshift=0pt]P-|O) -- ([yshift=0pt]P);
    \draw[mycoloroutput, line width=0.5pt] ([yshift=1.5pt]O) -- ([yshift=1.5pt]O-|P);
    \draw[mycoloroutput, line width=0.5pt] ([yshift=-1.5pt]P-|O) -- ([yshift=-1.5pt]P);
  },
  secondextra={
    \draw[mycoloroutput, line width=0.5pt] ([yshift=0pt]O) -- ([yshift=0pt]O-|P);
    \draw[mycoloroutput, line width=0.5pt] ([yshift=1.5pt]O) -- ([yshift=1.5pt]O-|P);
  },
  firstextra={
    \draw[mycoloroutput, line width=0.5pt] ([yshift=0pt]P-|O) -- ([yshift=0pt]P);
    \draw[mycoloroutput, line width=0.5pt] ([yshift=-1.5pt]P-|O) -- ([yshift=-1.5pt]P);
  }
]{coloredquotationoutput}

\PassOptionsToPackage{svgnames}{xcolor}
\usepackage{xspace, xpunctuate}
\usepackage{listings}  %
\usepackage{enumitem}  %
\usepackage{multirow}
\definecolor{keywordcolor}{rgb}{0.56, 0.13, 0.00}
\definecolor{ndkeywordcolor}{rgb}{0.05, 0.46, 0.17}
\definecolor{commentcolor}{rgb}{0.41, 0.64, 0.70}
\definecolor{stringcolor}{rgb}{0.25, 0.44, 0.63}
\lstdefinelanguage{TypeScript}{
  keywords={typeof, new, true, false, catch, function, return, null, catch, switch, var, if, in, while, do, else, case, break, boolean},
  morekeywords={[2]{class, export, throw, implements, import, this}},
  identifierstyle=\color{black},
  sensitive=false,
  comment=[l]{//},
  morecomment=[s]{/*}{*/},
  commentstyle=\color{commentcolor}\ttfamily,
  stringstyle=\color{stringcolor}\ttfamily,
  morestring=[b]',
  morestring=[b]"
}
\colorlet{punct}{red!60!black}
\definecolor{delim}{RGB}{20,105,176}
\colorlet{numb}{magenta!60!black}

\lstdefinelanguage{json}{
    basicstyle=\footnotesize\ttfamily,
    numbers=left,
    numberstyle=\color{gray}\footnotesize\ttfamily,
    numbersep=4pt,
    tabsize=2,
    showstringspaces=false,
    breaklines=true,
    literate=
     *{0}{{{\color{numb}0}}}{1}
      {1}{{{\color{numb}1}}}{1}
      {2}{{{\color{numb}2}}}{1}
      {3}{{{\color{numb}3}}}{1}
      {4}{{{\color{numb}4}}}{1}
      {5}{{{\color{numb}5}}}{1}
      {6}{{{\color{numb}6}}}{1}
      {7}{{{\color{numb}7}}}{1}
      {8}{{{\color{numb}8}}}{1}
      {9}{{{\color{numb}9}}}{1}
      {:}{{{\color{punct}{:}}}}{1}
      {,}{{{\color{punct}{,}}}}{1}
      {\{}{{{\color{delim}{\{}}}}{1}
      {\}}{{{\color{delim}{\}}}}}{1}
      {[}{{{\color{delim}{[}}}}{1}
      {]}{{{\color{delim}{]}}}}{1},
}

\lstdefinelanguage{mylang}{
}
\lstdefinestyle{mystyle}{
  basicstyle=\footnotesize\ttfamily,
  numbers=none,
  frame=none,
  columns=flexible,
  xleftmargin=0pt,
  aboveskip=0pt,
  belowskip=0pt,
  language=mylang
}

\newtcolorbox{bluequotation}{
  colback=black!10,  %
  sharp corners,  %
  boxrule=0.5pt,  %
  left=10pt,  %
  right=10pt,  %
}

\definecolor{sparkrectanglecolor}{rgb}{0.635,0.812,0.169}
\definecolor{sparkrectangle1color}{rgb}{0, 0, 0}

\definecolor{sparkspikecolor}{rgb}{0,0,0}

\definecolor{cartesian}{RGB}{165, 0, 38}
\definecolor{rectangle}{RGB}{255, 165, 0}

\newcommand{\oldtext}[1]{}

\newif\ifmarkfigure

\newcommand{\delete}[1]{\ignorespaces}
\newcommand{\revise}[1]{{\color{black}{#1}}}
\markfigurefalse

\newcommand{\vlbox}[1]{{\fcolorbox{gray!35}{gray!8}{\texttt{#1}}}}

\ifCLASSINFOpdf
  \graphicspath{{../pdf/}{../jpeg/}}
  \DeclareGraphicsExtensions{.pdf,.jpeg,.png}
\else
\fi

\begin{document}

\title{Visualization Generation with Large Language Models: An Evaluation}

\author{Xinyu Wang, Chenwei Liang, Shunyuan Zheng, Jinyuan Liang, Guozheng Li, Yu Zhang, \\ and Chi Harold Liu, \textit{Fellow}, IEEE %
\IEEEcompsocitemizethanks{
\IEEEcompsocthanksitem Xinyu Wang,  Chenwei Liang,  Shunyuan Zheng, Jinyuan Liang, Guozheng Li, Chi Harold Liu are with Beijing Institute of Technology. E-mail: \{ wang.xinyu, zsysoft, guozheng.li, chiliu\}@bit.edu.cn.
\IEEEcompsocthanksitem Yu Zhang is with the University of Oxford. Email: yuzhang94@outlook.com.
}%
}

\markboth{IEEE TRANSACTIONS ON VISUALIZATION AND COMPUTER GRAPHICS, December~2025}%
{Shell \MakeLowercase{\textit{et al.}}: A Sample Article Using IEEEtran.cls for IEEE Journals}

\IEEEpubid{}

\maketitle

\begin{abstract}
The frequent need for analysts to create visualizations to derive insights from data has driven extensive research into the generation of natural Language to Visualization (NL2VIS).
While recent progress in large language models (LLMs) suggests their potential to effectively support NL2VIS tasks, existing studies lack a systematic investigation into the performance of different LLMs under various prompt strategies.
This paper addresses this gap and contributes a crucial baseline evaluation of LLMs' capabilities in generating visualization specifications of NL2VIS tasks.
Our evaluation utilizes the nvBench dataset, employing \delete{GPT models} \revise{six representative LLMs and eight distinct prompt strategies to evaluate their performance in generating six target chart types using the Vega-Lite visualization specification.} 
\delete{The evaluation uses the nvBench dataset, employing both zero-shot and few-shot prompt strategies that cover the four categories identified in our taxonomy.}
\delete{Our experiment results demonstrate that GPT-3.5 and GPT-4 outperform existing NL2VIS techniques. 
Notably, few-shot prompts perform better than zero-shot prompts, with GPT-4 significantly outperforming GPT-3.5. 
Leveraging few-shot prompts based on GPT-4 yields the highest NL2VIS accuracy, reaching 62.80\%.}
\revise{We assess model performance with multiple metrics, including vis accuracy, validity and legality.}
\revise{Our results reveal substantial performance disparities across prompt strategies, chart types, and LLMs.}
\delete{summarize the limitations of GPT models} 
\revise{Furthermore, based on the evaluation results, we uncover several counterintuitive behaviors across these dimensions,} 
\revise{and propose directions for enhancing the NL2VIS benchmark to better support future NL2VIS research}.

\end{abstract}


\section{Introduction}

Data visualizations play an essential role in data analysis, helping in pattern identification and insight communication~\cite{munzner2014visualization}. 
However, creating compelling visualizations requires expertise in visualization design principles and proficiency in visualization authoring tools~\cite{Mackinlay-tg1986-autoGraphPre, mackinlay2007show}. 
One significant direction to make analysts concentrate on the data analysis itself is automating the creation of data visualization based on’ natural language queries from users.
Therefore, a surge of visualization-oriented natural language interfaces emerged, allowing data analysts to create visualizations simply by using natural language and promoting data analysis efficiency.

The development of natural language processing (NLP) techniques has greatly improved the feasibility of the NL2VIS approach. 
Many NL2VIS studies often use NLP libraries, such as Articulate~\cite{Sun-sg2010-Articulate} using Stanford Parser~\cite{StanfordParser}, DeepEye~\cite{Luo-sigmod2018-DeepEye} employing OpenNLP~\cite{OpenNLP}, and NL4DV~\cite{Narechania-tvcg2021-NL4DV} integrating CoreNLP~\cite{Manning-acl2014-CoreNLP}
\delete{To address these limitations, researchers have turned to deep learning-based methodologies} \revise{These methods, categorized as rule-based type, are constrained by their reliance on predefined rules, limiting the complexity of inputs and preventing the understanding of nuanced or complex natural language queries}~\cite{Shen-tvcg2023-NL2VISsvrvey}.
To address these limitations, researchers have adopted deep learning-based methodologies~\cite{Liu-pvis2021-ADVISor,Luo-tvcg2022-ncNet}, training neural networks to process more intricate natural language inputs.
However, \delete{a single} deep learning-based approaches \revise{utilizing conventionally-sized neural networks} often produce suboptimal performance and lack generalizability to perform diverse NL2VIS tasks.

Recently, there has been a proliferation of large language models (LLMs), exemplified by GPTs~\cite{OpenAI-arxiv230308774-GPT4report} showcasing their remarkable capability to comprehend natural language and generate responses in user-specified formats or programming \delete{codes} \revise{languages}~\cite{yang2023foundation}. 
These LLMs have demonstrated remarkable proficiency in a spectrum of generative tasks, encompassing code generation~\cite{hendrycksMeasuringCodingChallenge2021}, logical reasoning~\cite{liuEvaluatingLogicalReasoning2023}, and mathematical problem solving~\cite{yuanHowWellLarge2023}.
\revise{Moreover, the impressive capability of LLMs can be accessed with simple prompts, avoiding the need for users to conduct further time-consuming training.}
\delete{A plethora of research endeavors have evaluated the capabilities of LLMs from different aspects with diverse prompt strategies like Chain of Thoughts~\cite{weiChainofThoughtPromptingElicits}, Program of Thoughts~\cite{chenProgramThoughtsPrompting2022}, and Least to Most~\cite{zhouLeasttoMostPromptingEnables2023}.}

Therefore, it is also feasible to use LLM to perform NL2VIS tasks, leading to the development of several LLM-based systems, such as Chat2VIS~\cite{Maddigan-access2023-Chat2VIS} and LIDA~\cite{Dibia-acl2023-LIDA}, which generate Python codes to build data visualizations. 
In particular, Maddigan et al.~\cite{Maddigan-arxiv230314292-Chat2VIS} further evaluated the capability of the Chat2VIS system in NL2VIS tasks.

Beyond Python, domain-specific languages in JSON \delete{formats} \revise{grammar} have gained widespread adoption to specify visualizations in diverse applications~\cite{2021-grammar-chiea, McNutt2023Grammar}.
However, existing studies~\cite{Maddigan-arxiv230314292-Chat2VIS} have not evaluated the impact of various prompt strategies \delete{construction} on NL2VIS tasks.
Despite the substantial progress made in LLMs and LLM-based NL2VIS techniques, \delete{evaluating the capacity of LLMs using different prompt strategies in visualization generation is imperative} \revise{it is necessary to evaluate the capacity of LLMs using different prompt strategies in the visualization generation task}.
This evaluation serves as a foundational step in elucidating the practical utility of LLMs, driving enhancements, and ensuring responsible and informed deployment across various applications.

In this study, we aim to assess the capacity of LLMs to perform the NL2VIS task, focusing on utilizing the widely adopted Vega-Lite visualization grammar~\cite{Satyanarayan-tvcg2017-VegaLite} as the target output. 
To facilitate this evaluation, \revise{we employ eight different prompt strategies, tailored to meet the requirements of NL2VIS tasks.}
These prompts are structured to encompass three key components: role definitions for instructing LLMs, sampled data tables illustrating the data schema, and user queries delineating the desired visualization tasks encompassing expected chart types and relevant data attributes.
\delete{In crafting the zero-shot prompts, we incorporate static rules designed to instruct LLMs in revising errors identified in preliminary evaluations, thereby ensuring the generation of correct Vega-Lite specifications.
Conversely, for the few-shot prompts, we augment the prompt with multiple high-quality examples showcasing Vega-Lite specifications for diverse chart types.}
\delete{Each example comprises a query alongside its corresponding specification. 
We employ representative LLMs, namely GPT-3.5 and GPT-4, for the evaluation, leveraging the nvBench dataset~\cite{Luo-sigmod2021-NL2VISbenchmark} as the benchmark.}
We employ several representative open-source LLMs for the evaluation, using the nvBench dataset~\cite{Luo-sigmod2021-NL2VISbenchmark} as the reference.
Throughout the evaluation, we assess the performance of \delete{zero-shot and few-shot} \revise{each prompt strategy by computing the corresponding vis accuracy, validity and legality metrics.} \delete{by computing the matching accuracy between the visualization outputs generated by LLMs and the ground truth specifications provided in the nvBench dataset.}
\delete{This matching accuracy serves as a pivotal metric} \revise{These metrics serve as a pivotal role} in gauging the capabilities of LLMs for NL2VIS tasks, offering insights into their potential utility in this domain.

\delete{The evaluation results reveal a significant improvement in the performance of GPT models compared to previous studies in the task of Vega-Lite generation. 
The overall performance ranking of GPT models with different prompt strategies is as follows: GPT-4 (few-shot) $\textgreater$ GPT-4 (zero-shot) $\textgreater$ GPT-3.5 (few-shot) $\textgreater$ GPT-3.5 (zero-shot).
Notably, leveraging the GPT-4 model with few-shot prompts emerges as the most effective approach for data visualization creation, achieving an impressive accuracy rate of 62.80\%.  
Additionally, certain instances deviate from the overall performance ranking, particularly where GPT-3.5 with few-shot prompts outperforms GPT-4 with zero-shot prompts for specific chart types.
This discrepancy is attributed to the similar query statements in the benchmark dataset, which aid GPT-3.5 with few-shot prompts in effectively learning to handle similar queries.}

We conduct a detailed analysis of the evaluation results \revise{to uncover deep findings that can guide further studies} \delete{delve into the limitations of GPT models} and delve into the limitations of an existing benchmark. 
Despite the evident capabilities of \delete{GPT models} \revise{LLMs} in NL2VIS tasks, our evaluation results consistently reveal typical mistakes in the generated outputs. 
\delete{Upon a thorough examination, we identify three critical limitations in the capabilities of GPT models for NL2VIS tasks: inadequate comprehension of Vega-Lite syntax, Vega-Lite parameters, and given inputs.}
\revise{Upon detailed analysis, we identify three counterintuitive findings regarding the use of LLMs for NL2VIS tasks: more reasoning does not always lead to better performance, easier types do not always yield better results, and better-performing models do not always excel with every prompt strategy.}
Furthermore, beyond \delete{the limitations intrinsic to GPT models}\revise{these findings}, we found mistakes originating from the nvBench dataset benchmark itself, which require corrections to certain (NL, VIS) pairs. 
These errors in Benchmark span three primary categories: inaccuracies in query statements, improper data mapping, and incorrect underlying data/labels.
\delete{The identified capability deficiencies of GPT models negatively impact the performance of LLMs in NL2VIS tasks.}
\revise{Notably, these unexpected patterns, many of which were inconceivable prior to our evaluation highlight the way existing benchmark mistakes hinder further improvements in NL2VIS techniques.}
Importantly, this study uses these insights to deliberate on potential approaches to correct the benchmark and improve the efficacy of the NL2VIS techniques in future endeavors.

In summary, this paper has the following contributions:
\begin{itemize}[leftmargin=5mm]
    \item \textbf{Systematic Evaluation and Baseline.} We provide a comprehensive evaluation of LLMs' capacity for NL2VIS tasks, utilizing the nvBench dataset and Vega-Lite specifications under diverse prompt strategies, confirming the LLMs' ability to generate accurate specifications.
    \delete{ \item Evaluation of LLMs’ capacity for the NL2VIS tasks, employing Vega-Lite specifications with diverse prompt strategies.
    Our evaluation of LLMs utilizing the nvBench dataset underscores the capability of \revise{LLMs} to generate accurate Vega-Lite specifications.}
    \delete{ \item Identification of deficiencies in GPT models, including inadequate comprehension of Vega-Lite syntax, Vega-Lite parameters, and given inputs, serving as insights for enhancing NL2VIS techniques leveraging LLMs.}
    \item  \revise{\textbf{Novel Counterintuitive Findings.} We uncover three counterintuitive behaviors when using LLMs for NL2VIS: that easier chart types do not always yield better results, more reasoning is not always beneficial, and high-performing models do not uniformly excel across all prompts.}
    \item \textbf{Benchmark Validation and Improvement.} We identify critical limitations within the nvBench dataset, including incorrect query statements, improper visual mapping, and incorrect underlying data/labels. This analysis offers essential guidance for future work aimed at improving benchmark quality.
\end{itemize}

\section{Related Work}

We review the literature on the NL2VIS task and capability evaluation of LLMs to motivate our research.

\subsection{Natural Language to Visualization}

This section reviews existing studies on NL2VIS, classifying them into rule-based and \delete{neural network-based} \revise{deep learning-based} categories.
Since this work aims to evaluate the LLM capability for the NL2VIS task, we differentiate the LLM-based algorithms as the third category.

\textbf{Rule-based methods.}
The NL2VIS studies in the early stage always adopt rule-based algorithms to realize natural language comprehension and users' intent extraction.
Cox et al.~\cite{Cox-IJST2001-NLI2vis} propose a prototypical framework to use a natural language interface to generate data visualizations automatically.
However, the prototype system they developed exerts limitations on queries and can only generate data tables and bar charts.
Articulate~\cite{Sun-sg2010-Articulate} and DataTone~\cite{Gao-uist2015-DataTone} use the Stanford Parser library~\cite{StanfordParser} to analyze the natural language query and convert it to appropriate visualizations.
In particular, DataTone adds a customized widget to tackle the ambiguity that exists in the natural language query compared with Articulate.
Eviza~\cite{Setlur-uist2016-Eviza} adopts an ANTLR parser~\cite{Parr-spe1995-ANTLR}, strengthens the expressiveness of queries and can support users in interacting with data visualizations through natural languages.
DeepEye~\cite{Luo-sigmod2018-DeepEye} uses OpenNLP~\cite{OpenNLP} to parse an underspecified query consisting of keywords and rank multiple candidate visualizations to be selected.
FlowSense~\cite{Yu-tvcg2020-FlowSense} utilizes SEMPRE~\cite{Zhang-emnlp2017-SEMPRE} and CoreNLP~\cite{Manning-acl2014-CoreNLP} to perform semantic parsing and allows users to take advantage of dataflow visualization systems for data analysis by natural language interactions.
NL4DV~\cite{Narechania-tvcg2021-NL4DV} also adopts CoreNLP~\cite{Manning-acl2014-CoreNLP} to extract data attributes from the input query and decide the predefined tasks according to the query, and is interface-agnostic compared with previous approaches.
Although rule-based algorithms are relatively easy to apply, they have constraints on natural language inputs or cannot comprehend complex queries.
Thus, their performances are surpassed by subsequent \delete{neural network-based} \revise{deep learning} techniques. 

\textbf{\delete{Neural network-based} \revise{Deep Learning-based} methods.}
As deep learning techniques mature and become widely applied in NLP, the visualization community has begun to systematically explore \delete{neural network-based} \revise{deep learning-based} approaches to NL2VIS generation.
Early progress is made by Liu et al., who propose ADVISor~\cite{Liu-pvis2021-ADVISor}, a deep-learning-based system augmented with pre-defined rules to generate visualizations for tabular data.
Given a data table and a query in natural language, the system produces a visualization \delete{charts} with automated annotations that highlight user-specified information.
Luo et al.~\cite{Luo-sigmod2021-NL2VISbenchmark}, who present a synthesizer that leverages large-scale NL2SQL corpora to construct the nvBench benchmark, introduce a significant advancement in benchmarking and model training. This dataset, which contains roughly 25,000 pairs (NL, VIS) across more than 100 domains, provides a high-quality foundation validated through expert and crowd-sourcing evaluation. Building on nvBench, Luo et al. further develop ncNet~\cite{Luo-tvcg2022-ncNet}, a seq2seq model that maps natural language queries and datasets to Vega-Lite specifications, with optional chart templates enabling explicit chart-type control.
Beyond purely generative models, Song et al.~\cite{Song-kdd2022-RGVisNet} propose RGVisNet, which integrates retrieval-based and generation-based paradigms inspired by dialog systems and software development workflows.
\delete{The experiments show that the performance of RGVisNet surpasses previous NL2VIS approaches.}
Complementary to these approaches, Chen et al.~\cite{Chen-PL2022-synthesisNL2VIS} introduce Graphy, a system that combines program synthesis with BERT-based~\cite{Devlin-NAACL2019-BERT} NLP techniques. Their evaluation in NLVCorpus~\cite{Srinivasan-chi2021-nlvUtterance} demonstrates improved performance over previous rule-based and transformer-based solutions.
\revise{Additionally, Voigt et al.~\cite{voigt-etal-2024-plots} employ pre-trained language models fine-tuned on the nvBench~\cite{Luo-sigmod2021-NL2VISbenchmark} dataset, achieving state-of-the-art accuracy and near real-time performance in visualization generation. These findings further underscore the strong potential of leveraging language models for NL2VIS tasks and provide a valuable performance benchmark for our study.}

\textbf{LLM-based methods.}
With the growing interest in applying LLMs to NL2VIS, recent research has explored their strengths in generation, reasoning, and interaction for visualization tasks. 
Maddigan et al.~\cite{Maddigan-access2023-Chat2VIS} leverage LLMs to generate Python code for data visualization \revise{and subsequently extend their system to develop the Chat2VIS system with advanced features~\cite{Maddigan-arxiv230314292-Chat2VIS}, enabling multilingual inputs and iterative refinement of visualizations through extended conversational interactions.} 
\delete{They develop the Chat2VIS system to enable users to input a dataset and their analysis intentions through natural language, then the system converts queries to an appropriate prompt containing data table descriptions and queries in natural language and finally gives a desired chart rendered by generated Python code.}
LIDA~\cite{Dibia-acl2023-LIDA} conceptualizes \delete{\textit{automated visualization generation}} \revise{automated visualization generation} as a four-stage pipeline, integrating LLMs with image generation models to interpret datasets and analytic intents and ultimately produce both charts and infographics. 
\delete{The researcher introduces two metrics to evaluate the generated visualization: visualization error rate and self-evaluated visualization quality.}
Beyond chart generation, LLMs have also been leveraged to reduce the burden of data preparation and decision-making. Wang et al.~\cite{Wang-arxiv230910094-dataFormulator} introduce Data Formulator, enabling complex data transformations through natural language, while their subsequent work~\cite{Wang-arxiv231007652-LLM4Vis} employs ChatGPT for visualization recommendation accompanied by human-level explanatory rationales. 
Other efforts aim to strengthen the ecosystem around NL2VIS: Ko et al.~\cite{Ko-arxiv230910245-NLDatasetGenbyLLM} construct an LLM-based pipeline to generate high-quality natural language descriptions for real Vega-Lite charts, thereby enriching NL2VIS datasets. 
LLMs have further been adapted to more specialized visualization contexts. 
\revise{Yang et al.~\cite{acl24-Yang-MatPlotAgent} develop MatPlotAgent, an LLM-agent framework designed to improve scientific visualization generation, supported by a new benchmark, MatPlotBench, for comprehensive evaluation.} 
\revise{Tian et al.~\cite{tvcg25-Tian-ChartGPT} propose ChartGPT, which adopts the Least-to-Most prompt strategy~\cite{zhouLeasttoMostPromptingEnables2023} to guide LLMs through step-by-step visualization generation, yielding improved transformation quality.}

The LLMs-based NL2VIS task has recently attracted attention. 
However, current research predominantly focuses on LLM-based methods for Python code generation, leaving a gap in exploring visualization generation techniques grounded in a declarative visualization grammar. 
Moreover, there is a distinct lack of investigation into the efficiency of various prompt strategies for this task, and consequently, a conspicuous absence of a performance baseline for LLMs in the NL2VIS context. 
This paper aims to address these critical gaps and enhance the comprehension of these underexplored aspects.

\subsection{Capability Evaluation of LLMs}

LLMs have recently showcased remarkable capabilities across a spectrum of natural language processing tasks.
However, the quantitative performance of LLMs in different specific tasks is unclear, which may be unbeneficial for users to take advantage of LLMs.
Assessing these models' actual quality, capabilities, and limitations is necessary to assist users in utilizing LLMs more effectively.
Previous studies have systematically evaluated the capabilities of LLMs from different perspectives, such as code generation, reasoning, and mathematics, but there are still gaps to be filled in the visualization task. 

\textbf{Code generation.} 
Converting natural language to code usually includes automatic code generation, translation, and code completion.
The generated code needs to meet multiple standards, such as functional requirements, grammatical correctness, and coding style, making the evaluation a complex task. 
Hendrycks et al.~\cite{hendrycksMeasuringCodingChallenge2021} introduce APPS, a code generation benchmark to systematically evaluate the capability of LLMs to generate Python code directly from natural language specifications. 
Moreover, Cassano et al.~\cite{cassanoMultiPLEScalableExtensible2022} build the first multilingual code generation benchmark, MultiPL-E, providing significant experience for understanding and improving multilingual language models in code generation tasks.
Furthermore, Liu et al.~\cite{liuYourCodeGenerated2023} propose a framework to expand the evaluation dataset.
The framework uses a generator to obtain a considerable test set to cover different code paths and edge cases, thereby rigorously evaluating the accuracy of LLM-generated code.
Based on the benchmark, Ding et al.~\cite{dingStaticEvaluationCode2023} present a static code-completion evaluation framework. 
It analyzes the errors made by pre-trained LLMs on a real-world Python evaluation set and identifies common static errors and the trends of their occurrence frequency.

\textbf{Visualization.}
Evaluating the capabilities of LLMs for visualization \delete{generation} tasks is crucial because the evaluation results help users understand the performance of LLMs and guide further improvements and optimizations.
\revise{In addition to the LLM-based NL2VIS methods introduced in the previous part, some work focuses on the evaluation of LLMs' capability for the visualization generation task.}
Maddigan et al.~\cite{Maddigan-arxiv230314292-Chat2VIS} quantitatively evaluate the performance of the proposed Chat2VIS system~\cite{Maddigan-access2023-Chat2VIS} based on Python using benchmark datasets, and the results show that their system has a performance comparable to previous NL2VIS approaches.
\delete{However, Chat2VIS outputs Python codes, which is not the mainstream approach for visualization creation.
Moreover, they have not fully optimized their prompts with various existing strategies and the performance cannot present the true potential of LLMs in the NL2VIS task.}
\revise{Vazquez~\cite{pacificvis24-Vazquez-LLMready4vis} evaluates the capabilities of LLMs to generate visualizations of 24 different charts and three different Python libraries using GPT3 and GPT4.
The results show that LLMs can support a majority of tested chart types  regardless of specific Python libraries.}
\revise{Furthermore, some work proposes new frameworks to provide guidance for the NL2VIS evaluation.}
\revise{Podo et al.~\cite{eurova24ws-Podo-framework4eval} propose a theoretical framework, EvaLLM, for the evaluation of LLM-based NL2VIS approach, which lists multiple visualization properties to be assessed.}
\revise{Chen et al.~\cite{tvcg25-Chen-VisEval} propose VisEval, a high-quality benchmark for LLM-based NL2VIS evaluation, and an evaluation framework that has three main perspectives: validity, legality, and readability.}

\revise{In addition to the visualization generation task, some work is interested in the capability of LLMs for other tasks.}
\revise{Kim et al.~\cite{arxiv231009617-Kim-HowGoodonVisDesign} evaluate the capability of ChatGPT to generate guidance on visualization design and find that ChatGPT has great visualization knowledge and can give various responses compared with human experts.}
\revise{Chen et al.~\cite{arxiv230602914-Chen-BeyondGeneratingCode} evaluate the capability of GPT models to accomplish the assignments of a data visualization course and find the potentials of GPT to complete more visualization tasks.}
\revise{Wu et al.~\cite{emnlp24-Wu-ChartInsights} focus on the capability of multimodal LLMs for chart question answering tasks.
They make a new dataset, ChartInsights, and evaluate the performance of 19 multimodal LLMs.}

In this paper, we investigate existing prompt strategies and \delete{leverage} \revise{apply} the strategies \revise{to make them} suitable for our NL2VIS task to construct different high-quality prompts.
We selected Vega-Lite~\cite{Satyanarayan-tvcg2017-VegaLite}, which is a popular approach for the creation of visualization in the visualization community, and we evaluated the performance of LLMs \delete{using few-shot and zero-shot prompts} to explore the potential of LLMs in the NL2VIS task \revise{under different prompt strategies}.

In summary, while extensive studies have been carried out on LLMs evaluation, a significant gap still exists in the comprehensive examination of the visualization task \delete{construction} \revise{generation} \revise{for LLMs under different prompt strategies}. 
This work aims to evaluate the performance of the NL2VIS task from the declarative grammar aspect and to establish a foundation to improve the efficiency of LLM-based visualization generation.

\section{Prompts Design}
In this section, we first \delete{summarize} \revise{review} the prompt strategies and \delete{analyze whether a strategy is suitable for the NL2VIS task} \revise{introduce the strategies we adopt in our NL2VIS evaluation}.
Then, we introduce the \delete{evaluated} \revise{constructed} prompts and \delete{the iterative process of these prompts} \revise{the process that we apply different strategies to the NL2VIS task}.

\begin{figure}
    \centering
    \includegraphics[width=\columnwidth]{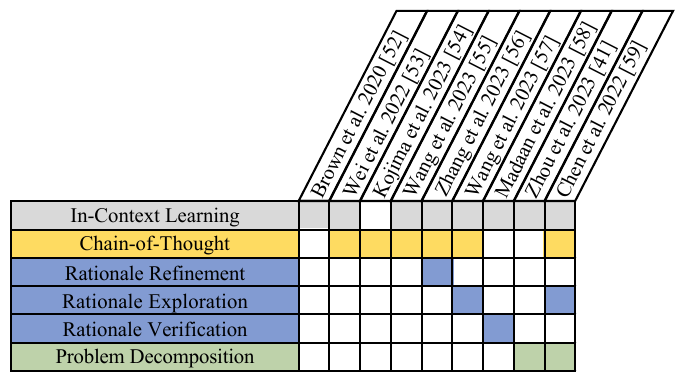}
    \caption{
        \textbf{A summary of prompt strategies:}
        \revise{Each column corresponds to a study that utilizes or evaluates prompt strategies.
        Each row corresponds to a prompt strategy category.
        The strategies can be grouped into four categories: In-Context Learning (gray), Chain-of-Thought and its variants (yellow), Rationale Engineering (blue) and Problem Decomposition (green).}
    }
    \label{fig:prompt_strategy}
\end{figure}

\subsection{Prompt Strategies}
\label{sec:prompt-strategies}

A prompt strategy involves crafting effective prompts or instructions to interact with LLMs to elicit a desired response. 
Reasonable design and selection of these prompts are crucial, as they enable task customization, facilitate control over model output, aid in preventing misunderstandings, and contribute to resource efficiency.

\delete{This section divides the commonly used prompt technologies that promote LLMs reasoning for specific tasks or settings into four categories (see Fig.~\ref{fig:prompt_strategy}) and then analyzes whether each category is suitable for visualization generation tasks.}
\revise{We leverage the categorization proposed by Huang et al.~\cite{acl23-Huang-TowardsReasoningLLMSurvey} and divide commonly used prompt strategies into four categories (see Fig.~\ref{fig:prompt_strategy}): In-Context Learning, Chain-of-Thought and Its Variant, Rationale Engineering and Problem Decomposition.
For each type, we introduce typical prompt strategies we adopt in the evaluation.}
\delete{It is worth emphasizing that different prompt strategies can be combined, and multi-stage prompting processes can be integrated, but it does not always need to be.}

\subsubsection{In-Context Learning (ICL)}
ICL can improve the predictive performance of LLMs by improving contextual information through examples or instructions relevant to the task.
\revise{The few-shot prompt is a typical ICL strategy~\cite{nips20-Brown-FewShotLearners}, for which the LLM receives examples described in natural language templates and its demonstration contexts as prefixes, followed by a query as input, thus generating outputs that imitate the provided examples.
Furthermore, considering that the zero-shot prompt having no demonstration examples can also be added some contextual information of tasks, we regard the zero-shot prompt as another ICL strategy and adopt it in our evaluation.}
\delete{For the ICL prompting strategies, the LLM receives examples described in natural language templates and its demonstration contexts as prefixes, followed by a query as input, thus generating outputs that imitate the provided examples.}
\delete{
ICL can be combined with other prompting techniques like Chain-of-Thought, with the demonstration context encompassing the process of reasoning, not just the answers. 
This approach can effectively stimulate the reasoning ability of LLMs.
}
\delete{Yin et al.~\cite{yinNaturalLanguageCode2023} employ a prompt including notebook contexts and the current intent into NL2Code tasks, making LLMs directly comprehend the task based on provided instances.}

\subsubsection{Chain-of-Thought (CoT) and Its Variant}
In \delete{standard} \revise{the basic} prompt, LLMs are always asked to solve intricate, multi-step problems in a singular step, while the human cognitive methodology involves tackling complex reasoning problems incrementally.
To address this, Wei et al.~\cite{weiChainofThoughtPromptingElicits} introduce the CoT methodology, which explicitly integrates multiple reasoning path steps before generating the ultimate answer.
This allows LLMs to showcase their intermediate reasoning process, leading to more accurate and explainable results.

Using a few-shot approach, CoT requires a manual effort to design prompts.
Kojima et al.~\cite{kojimaLargeLanguageModels2023a} demonstrate that LLMs can function as zero-shot reasoners by simply appending the instruction ``Let’s think step by step'' before the response. 
\delete{We also explore some step-by-step prompting strategies in the preliminary study and find that they have no significant improvements in performance. 
Therefore, we add several rules about the Vega-Lite grammar to construct zero-shot prompts instead.}
Additionally, Wang et al.~\cite{wangPlanandSolvePromptingImproving2023} propose the Plan-and-Solve prompt. This enhanced zero-shot CoT strategy first requires the model to formulate an initial plan to systematically break down the overarching task into smaller, manageable sub-tasks, and then to execute them according to the devised plan.
In our evaluation, we adopt both of these CoT-related strategies.

\subsubsection{Rationale Engineering}
Rationale Engineering is a set of techniques designed to leverage the LLM's intrinsic reasoning capabilities more effectively than the original CoT prompting. 
While the original CoT prompt relies on artificially crafted examples of intermediate reasoning steps and employs greedy decoding to generate a single coherent reasoning process, rational Engineering focuses on optimizing, exploring, and verifying the reasoning chains.

\textbf{Rationale refinement} focuses on optimizing the quality of the input reasoning steps. It involves providing models with more examples of intermediate steps or optimizing the structure and expression of these examples.
\revise{For example, Zhang et al.~\cite{iclr23-Zhang-autoCoT} propose Auto-CoT that can automatically generate CoT examples.
Auto-CoT selects typical questions from a related dataset by clustering and use Zero-shot-CoT~\cite{kojimaLargeLanguageModels2023a} to generate reasoning processes for the questions, thereby constructing CoT examples.}
\delete{
Fu et al.~\cite{fuComplexityBasedPromptingMultiStep2023} propose complexity-based consistency, suggesting that prompts with higher reasoning complexity have higher performance on multi-step reasoning tasks.
Lu et al.~\cite{luDynamicPromptLearning2023} propose PROMPTPG, employing policy gradient to strategically select in-context examples from a subset of training datasets and then construct corresponding prompts for the selected ones.
}
\delete{We specifically select diverse examples in the few-shot prompt, covering all categories of charts in various difficulty levels, to improve the performance of LLMs.}

\textbf{Rationale exploration} aims to select the best reasoning chain by discovering and evaluating multiple potential reasoning paths. 
Wang et al.~\cite{wangSelfConsistencyImprovesChain2022} propose a self-consistency decoding strategy, which involves generating multiple independent inference paths and then performing majority voting among them to obtain the final, most rebust answer.
\delete{However, the intermediate results of the NL2VIS task in this work are Vega-Lite specifications, with a single visualization image potentially corresponding to multiple Vega-Lite specifications.
Measuring the similarity between Vega-Lite codes is challenging, making the rationale exploration approach not applicable to our work.}

\textbf{Rationale verification} can be introduced to verify the validity and rationality of the reasoning steps after generating the reasoning paths.
\delete{Weng et al.~\cite{wengLargeLanguageModels2023} propose a self-verification to enable LLMs with the capability of verifying answers.
A similar prompt ensembling strategy includes DIVERSE~\cite{liMakingLargeLanguage2023}, a reasoning step validator to filter out incorrect answers and independently verify each inference step.}
\revise{Madaan et al.~\cite{nips23-Madaan-SelfRefine} propose a method the LLM is prompted to give feedback and refine its own output iteratively. This technique allows the LLM to generate high-quality responses without human assistance by ensuring self-correction and quality control within the reasoning process.}
\delete{We do not adopt the rationale verification strategy because LLMs can only output results in textual form when evaluations and the validity and correctness of a Vega-Lite specification cannot be verified without the corresponding visualization image.}

\subsubsection{Problem Decomposition}
Complex tasks often require systematic analysis and solution development.
One common strategy for managing complexity is problem decomposition, which is designed to break down a large problem into smaller, more manageable sub-problems. 
This approach helps prevent LLMs from being overwhelmed by the task's scope.
Problem decomposition strategies can be divided into two types.
Zhou et al.~\cite{zhouLeasttoMostPromptingEnables2023} present the Least-to-Most prompt, which deconstructs a complex problem into a sequence of multiple sub-problems.
Critically, the solutions derived from earlier sub-problems are then utilized as prompts to inform and address the subsequent ones, building towards the final answer.

\revise{For each prompt type, there exist strategies that are not suitable for the NL2VIS task, thus we do not adopt them in our evaluation.
For instance, the Program of Thoughts (PoT) prompt strategy introduced by Chen et al.~\cite{chenProgramThoughtsPrompting2022}, which leverages LLMs pre-trained with codes to script programs and decouples calculations from reasoning tasks to promote the correctness of the calculation process.
Since our NL2VIS task using Vega-Lite is primarily a code generation task without requiring computational reasoning, the PoT strategy is not applicable in our case.}
\subsection{Evaluated Prompts}
\label{sec:evaluated-prompts}
\delete{
The LLMs can take text as input and generate codes.
Therefore, it is a promising approach to finish the NL2VIS tasks based on LLMs.
Because existing NL2VIS techniques are end-to-end, that is, they accept natural language queries and output visualization specifications, 
}
\revise{This section outlines the methods we use to apply different prompt strategies to the NL2VIS task and provides specific prompt examples.}
For this work, we focus solely on the one-round interaction condition to simulate an end-to-end paradigm and simplify the evaluation process.
We define the LLM-based NL2VIS process as two steps.
First, users input a prompt that contains task descriptions and related data tables. 
Second, the LLMs return the Vega-Lite specifications as a response, which are then used to produce the corresponding visualization results.

To make LLMs \delete{(in this paper, GPT-3.5 and GPT-4)} generate correct and effective visualizations, we optimize our prompts and compare the respective performances on the nvBench dataset.
The basic version of out prompt is straightforward: it merely instructs the LLM on its roles in the system message, the task query, and the data table it should use in the user message, as presented below.

\begin{bluequotation}
\begin{lstlisting}[style=mystyle, breaklines=true, keepspaces=true]
{
    "role": "system",
    "content": "You are a data analysis assistant that uses Vega-Lite to create data visualizations."
}
\end{lstlisting}
\end{bluequotation}
\begin{bluequotation}
\begin{lstlisting}[style=mystyle, breaklines=true, keepspaces=true]
{
    "role": "user",
    "content": f"Create visualization using Vega-Lite that has correct syntax and uses appropriate visualization type for the {database_name} dataset to complete this task:
    ```{nl_query}```
    The sampled {database_name} database is as follows:
    ```{sampled_data}```
    The "data" attribute of your vega-lite json must be: "data": {{"url": "data.csv"}}."
}
\end{lstlisting}
\end{bluequotation}

The variable in the curly brackets \textit{database\_name}, \textit{nl\_query}, and \textit{sampled\_data} are the name of the database, the task description, and the needed data table extracted from the nvBench dataset, respectively.
Since several data tables in the nvBench are very large, we transfer a sampled data table to \delete{GPT models} \revise{LLMs} to decrease the number of tokens.
\delete{We find that the output given by GPT models is not a pure Vega-Lite specification and always contains redundant explanatory text.}
\delete{To make the output only have Vega-Lite specification \delete{to facilitate following automatic processes}, we add an instruction at the end of the prompt to make GPT models only output the Vega-Lite specification.}
\delete{The generated specifications always use inline data sources in the Vega-Lite specification, which might cause errors because the prompt only consists of a partial data table sampled from the original data.}
\delete{Therefore,}
\revise{To standardize the format of the \vlbox{data} property in the specification, we instruct the LLM to use the URL ``\textit{data.csv}'' as the data source. The URL is replaced by the needed data table before the visualization generation.}
\revise{Additionally, for LLMs that do not support system prompts, we merge the system prompt with the user prompt.}

We use the basic version of the prompt as the foundation for constructing the subsequent prompt variations, with full templates and implementation details provided in the Appendix A. \delete{zero-shot and few-shot}

\subsubsection{Zero-shot}
The Zero-shot prompt provides the model with task descriptions, a sampled data table, and a concise set of generation rules. 
These rules are specifically designed to prevent common specification errors observed in a preliminary analysis of nvBench instances. 
Based on our preliminary analysis, we identified several recurring issues in the Vega-Lite specifications generated by the model and incorporated five targeted rules to mitigate them.
First, the prompt explicitly requires the specification to adopt Vega-Lite v5 to avoid inconsistencies caused by mixed version outputs. 
Second, it enforces the correct ordering between the \vlbox{transform} and \vlbox{encoding} blocks, ensuring that the fields created in the transformation steps can be referenced correctly. 
Third, the prompt reminds the model to include the necessary \vlbox{filter} operations when the query requires data selection. 
Fourth, it instructs the model to specify the \vlbox{groupby} field whenever aggregation is applied, preventing incorrect default behaviors. 
Fifth, it clarifies the proper placement of sorting instructions outside the \vlbox{transform} object to align with the Vega-Lite syntax.
These rules collectively guide the model toward producing syntactically valid and semantically faithful visualizations. The design purpose of this strategy is to provide essential structural guidance that reduces systematic specification errors without imposing task-specific demonstrations while maintaining a minimal zero-shot setting that avoids additional examples or external constraints.

\subsubsection{Few-shot}
The few-shot prompt strategy incorporates a curated set of examples composed of task queries paired with their corresponding correct Vega-Lite specifications. 
These examples are selected from the nvBench dataset and the specifications are derived from previously validated outputs. 
To ensure broad task coverage, we construct an example for each chart type defined in Vega-Lite, thereby avoiding performance degradation caused by missing chart categories. 
Each example is designed to contain at least one representative operation, such as \vlbox{sort}, \vlbox{filter}, or \vlbox{aggregate}, allowing the model to observe diverse transformation patterns. Through these exemplars, LLMs are expected to acquire grammar regularities of the Vega-Lite grammar and generalize the learned behavior to unseen queries. 
The purpose of this strategy is to provide grounded demonstrations that improve the adherence to domain-specific conventions and reduce grammar inconsistencies in the generated specifications.

\revise{
\subsubsection{Zero-shot-Chain-of-Thought (Zero-shot-CoT)}
The Zero-shot-CoT strategy proposed by Kojima et al.~\cite{kojimaLargeLanguageModels2023a} simply appending the phrase ``Let’s think step by step'' to the prompt to encourage structured intermediate reasoning. 
In the NL2VIS scenario, we place this instruction immediately after the sampled data table and before the explicit constraint on the \texttt{data} property required for Vega-Lite specifications. 
This positioning serves to anchors the reasoning process to the observed dataset while encouraging stepwise deliberation prior to generating syntactically valid Vega-Lite specification. 
The strategy may optionally include a system-level prompt that frames the model as a Vega-Lite data analysis assistant, reinforcing task alignment. 
The design intention is to strengthen dataset-aware reasoning and improve the consistency of the resulting specifications.

\subsubsection{Plan-and-Solve Plus (PS-Plus-CoT)}
The PS-Plus-CoT prompt strategy~\cite{wangPlanandSolvePromptingImproving2023} enhances structured reasoning by explicitly guiding the model to first interpret the problem, identify key elements, formulate a plan, and then execute that plan step by step. 
For NL2VIS, we adapt this instruction to emphasize reasoning steps specific to Vega-Lite specification generation: the model is directed to interpret the query, identify relevant data operations, such as \vlbox{filter} and \vlbox{aggregate}, recognize visualization decisions related to \vlbox{encoding} and \vlbox{sort}, and finally produce the Vega-Lite specification while maintaining grammatical correctness. 
The PS-Plus-CoT instruction is inserted at the same location as the Zero-shot-CoT suffix (i.e., after the data table). 
This design aims to provide an explicit procedural scaffold that improves coherence in the generation of complete visualization specifications.

\subsubsection{Auto-Chain-of-Thought (Auto-CoT)}
Auto-CoT~\cite{iclr23-Zhang-autoCoT} automatically constructs CoT demonstrations, thereby reducing the reliance on manual prompt engineering. In our implementation, Auto-CoT extends the Zero-shot-CoT template by adding automatically generated CoT examples to the beginning of the prompt. 
To select representative queries from nvBench, all training queries are clustered using K-Means (with the number of clusters set to three).
One representative query from each cluster is used to elicit a CoT-style rationale and its associated Vega-Lite specification under the Zero-shot-CoT prompt. 
These examples are formatted into a unified demonstration block and appended to the main prompt.
The purpose of this design is to provide task-specific exemplars that reflect the model’s own reasoning tendencies, thereby leveraging the benefits of few-shot learning without human intervention.

\subsubsection{Least-to-Most}
The Least-to-Most prompt strategy~\cite{zhouLeasttoMostPromptingEnables2023} decomposes complex reasoning tasks into a sequence of simpler subtasks. 
Following the decomposition framework of Tian et al.~\cite{tvcg25-Tian-ChartGPT}, we reformulate the NL2VIS pipeline as a series of intermediate stages, including schema initialization, data specification, column selection, optional application of \vlbox{filter} and \vlbox{aggregate} operations, chart type determination, \vlbox{encoding} selection, application of \vlbox{sort} rules, and final assembly of the Vega-Lite specification. 
An illustrative example from the nvBench training set is appended to the base prompt to construct the complete Least-to-Most template. 
This design encourages the model to follow an ordered reasoning trajectory consistent with the inherent multi-stage structure of visualization generation.

\subsubsection{Self-Refine}
The Self-Refine prompt strategy~\cite{nips23-Madaan-SelfRefine} enables LLMs to iteratively critique and improve their own generated Vega-Lite specifications. 
For each query, the model first produces an initial specification, then analyzes it for potential issues, including incorrect \vlbox{encoding}, inappropriate transformations, or missing \vlbox{filter} operations. 
Based on the detected issues, it produces a refinement.
Three exemplar cases, each consisting of a query, an initial specification, and identified problems, are provided to demonstrate the intended refinement process.
The iteration count is capped at three to avoid infinite loops, and a standardized message is used when no refinement is needed. 
This design promotes progressive improvement, enhances syntactic correctness, and increases the overall robustness of NL2VIS generation.

\subsubsection{Self-Consistency}
The Self-Consistency strategy~\cite{wangSelfConsistencyImprovesChain2022} identifies the most reliable output by selecting the answer that appears most frequently across multiple generations. 
In NL2VIS, although different outputs may produce equivalent visualizations, their textual Vega-Lite specifications often differ. 
To address this, we extract the data attributes from each generated visualization and group outputs that share the same attributes. 
The final result is sampled from the largest group, ensuring that the selected output reflects the most consistent data representation. 
While this method improves robustness by leveraging diversity across generations, it exhibits relatively low computational efficiency because it requires collecting and comparing multiple outputs before a consensus can be established. The design intention is to promote consistency and mitigate the influence of outlier generations in visualization tasks.}

\section{Evaluation}
This section will first introduce the dataset and \delete{LLM models} \revise{LLMs} utilized in our evaluation. 
Subsequently, we present and analyze the initial evaluation results. \delete{, comparing them with previous approaches in the NL2VIS domain.}
\revise{Detailed findings summarized from the results will be provided in the next section.}

\subsection{Dataset and Model}

We \delete{employed} \revise{utilize} the nvBench~\cite{Luo-sigmod2021-NL2VISbenchmark} datasets as the benchmark for our evaluation.
The nvBench dataset comprises more than 7,247 visualization instances covering various application scenarios such as sports and medical.
Each instance in the nvBench comprises a task described in various natural language queries, input data tables, and the ground truth, which includes a Vega-Lite specification and a visualization image.
The nvBench dataset categorizes NL2VIS tasks into four difficulty levels: easy, medium, hard, and extra hard.
The ground truth visualizations in this dataset cover seven chart types: bar, pie, scatter, stacked bar, \revise{line, grouping line}, and grouping scatter charts.

A notable observation within the nvBench dataset is that certain instances require joining at least two data tables to fulfill the associated task. 
However, Vega-Lite lacks native support for incorporating multiple data sources into a single visualization specification.
Consequently, to address this constraint, it is required to merge multiple data tables, filter the requisite attributes using SQL queries, and utilize the resulting data processing outcome as input for the Vega-Lite specification. 
Considering that this study exclusively evaluates the capability of LLMs to generate visualizations in the form of Vega-Lite specifications, we filter out instances related to multiple data tables that necessitate this kind of multi-table preprocessing based on SQL, and focus on evaluating the remaining instances.
\revise{Furthermore, upon inspecting the line charts and grouping line charts in nvBench, we discovered that a part of the given ground truths aggregate data over multiple years into one time unit on the x-axis even the corresponding queries do not require such aggregation.
This inherent mismatch between the LLM-generated visualizations and the nvBench ground truths would skew the evaluation.
Consequently, we filter out these instances to exclude all grouping line charts from the final evaluation set.
Thus, the obtained evaluation statistics do not contain the results of grouping line charts.}

In terms of LLMs, we select \delete{various} \revise{six} representative \revise{open-source} LLMs\revise{ with small parameter numbers: QWEN2-7B-Instruct (QWEN2)~\cite{qwen2}, 
Meta-Llama-3-8B-Instruct (Llama3)~\cite{llama3modelcard}, Mistral-7B-Instruct-v0.3 (Mistral)~\cite{jiang2023mistral7b}, Gemma-7b-it (Gemma)~\cite{gemma_2024}, GLM-4-9B-Chat (GLM V4)~\cite{glm2024chatglm}, and DeepSeek-R1-Distill-Llama-8B (Deepseek-R1-llama3.1)~\cite{deepseekai2025deepseekr1incentivizingreasoningcapability}, to implement the evaluation.} \delete{GPT-3.5 and GPT-4  as representatives due to their widespread adoption.}
\delete{To handle prompts containing few-shot examples and sampled data tables without surpassing token limitations, we employ the GPT-3.5-turbo-16k and GPT-4 API for implementation.}
We configure the temperature parameter of the \delete{GPT models} \revise{used LLMs} to zero to reduce randomness during generation.
\delete{In particular, given the high computational cost associated with generating specifications using GPT-4, we randomly sample 1,000 instances from the nvBench dataset to evaluate the performance of GPT-4 in NL2VIS tasks.}
\revise{Furthermore, as the previous NL2VIS related work utilizes the test set of dataset in the evaluation stage~\cite{Luo-tvcg2022-ncNet}, we also utilize the test set of the nvBench in the evaluation.}

\subsection{Metrics}
\label{sec:metrics}

\revise{
To evaluate the capability of LLMs to generate Vega-Lite specifications for the NL2VIS task, we employ three evaluation metrics: vis accuracy\cite{Luo-sigmod2021-NL2VISbenchmark}, validity~\cite{tvcg25-Chen-VisEval} and legality~\cite{tvcg25-Chen-VisEval}.
The vis accuracy evaluates whether the chart type of the generated visualization matches the ground truth.
Given that different chart types have distinct \vlbox{mark} properties in the Vega-Lite specification, we evaluate performance by comparing the generated and ground truth \vlbox{mark} values.
We calculate the ratio of matching items to the total number of cases as the vis accuracy.
Validity measures whether the Vega-Lite specification can be successfully rendered into a visualization.
Specifications that violate the Vega-Lite grammar are considered invalid. 
We compute the validity as the percentage of valid specifications.

Legality measures the capacity of a Vega-Lite specification to fulfill the task described in the query. 
For our evaluation, we compare the data attributes contained in the generated visualizations (in SVG format) against the ground truth visualization. 
Specifically, a generated specification is considered legal only if the data attributes of both visualizations are identical. 
The final legality score is the percentage of specifications deemed legal.
}
The ground truths provided in the nvBench dataset comprise both a Vega-Lite specification and the corresponding visualization image. However, the Vega-Lite specification ground truths utilize the transformed data rather than raw data from the input table, and lack the data transformation definitions in the specification.
This absence makes it challenging to directly compare the specification in the ground truth with those generated by \delete{GPT models} \revise{LLMs}. 
Therefore, in our evaluation, we focus on the \revise{data} matching \delete{accuracy} between the visualization ground truth and the visualization generated by \delete{GPT models} \revise{LLMs to determine the legality.}
\delete{To achieve this, we convert the generated Vega-Lite specifications into visualizations and then compare them with the ground truth visualizations to determine their equivalence.}
\revise{Furthermore, this visualization-based approach is broadly applicable to calculating legality for any visualization grammar that produces output in the SVG format.}

\subsection{Evaluation Results}
\label{subsection:results}

This section presents the evaluation results of \delete{GPT-3.5} \revise{LLMs} \delete{models} in the nvBench dataset. 
Additionally, we conduct a comparative analysis of the performance \delete{between the GPT-3.5 and GPT-4 models} of different LLMs \delete{models} using \delete{sampled instances from} \revise{the test set of} the nvBench dataset.
\revise{Table~\ref{tab:total-results} summarizes the average performance of all evaluated LLMs across different prompt strategies, including the validity rate, legality rate, and visualization accuracy. Table~\ref{tab:total-legality} further reports the overall legality outcomes stratified by LLMs, chart types, and task difficulties.
We present only this representative detailed table in the main text; the complete set of detailed legality results for all models is provided in the Appendix B.
A more in-depth analysis and interpretation of these results is presented in Section~\ref{subsection:deep-findings}.
}

\subsubsection{Performance of Different Prompt Strategies}
\label{sssec:prompt}
\begin{figure}
    \centering
    \includegraphics[width=\columnwidth]{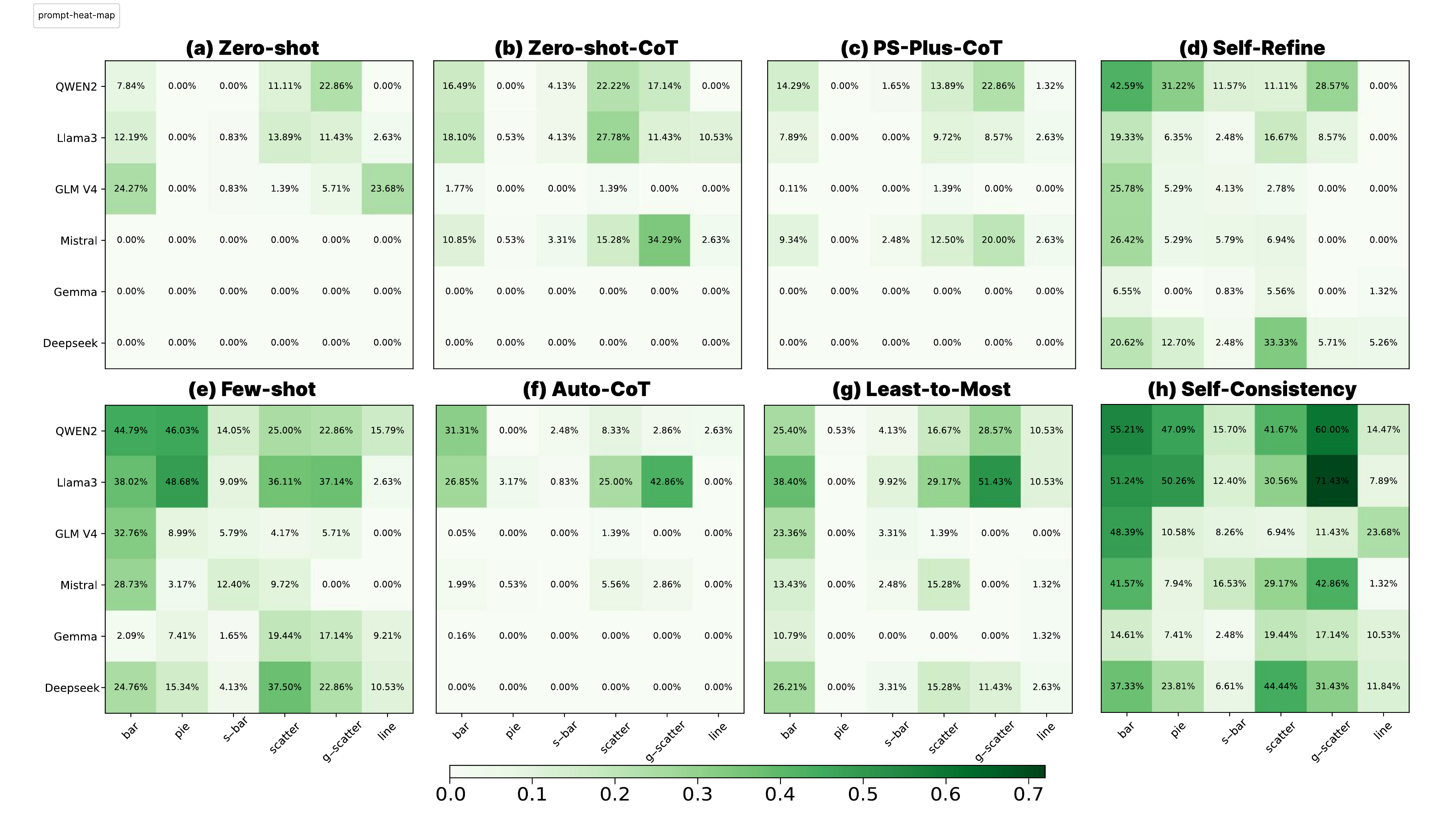}
    \caption{
        \textbf{Legality heat map of prompts} 
        The figure presents the legality performance heat map of different prompt strategies.
    }
    \label{fig:prompt-heat-map}
\end{figure}

\revise{
From Table~\ref{tab:total-results}, the prompt strategies exert a substantial influence on both the valid rate and the legal rate. 
In contrast, the visualization accuracy remains consistently high, between 95\% and 97\%, indicating that LLMs can accurately identify the intended chart type regardless of the prompt strategy. 
The valid rate and legal rate exhibit similar ranking patterns, where a higher valid rate generally corresponds to a higher legal rate. Among all strategies, the Self-Consistency prompt achieves the best performance, while the Zero-shot-CoT prompt yields the poorest results. The approximate performance ranking is: Self-Consistency\textgreater Few-shot\textgreater Least-to-Most, Self-Refine\textgreater Auto-CoT, Zero-shot, Zero-shot-CoT\textgreater PS-Plus-CoT. 

Beyond the aggregated statistics in Table~\ref{tab:total-results}, Fig.~\ref{fig:prompt-heat-map} provides further insights by illustrating the performance variation of each prompt strategy across different LLM–chart-type combinations. Prompt strategies with similar overall scores exhibit nearly identical heat map patterns, implying that their strengths and weaknesses are consistent across evaluation conditions. Importantly, prompting does not fundamentally alter which LLM–chart-type combinations perform well or weakly: combinations with high legality remain strong across all strategies, while low-legality combinations persistently underperform. 
}

\delete{Table~\ref{tab:results} presents the evaluation results using the SVG-based matching strategy.
The overall matching accuracies of zero-shot and few-shot prompt strategies on the nvBench dataset are 43.23\% and 50.12\%, respectively, with the performance of the few-shot prompt significantly outperforming that of the zero-shot prompt.
Concerning performance across different chart types, the zero-shot prompt achieves the highest accuracy with scatter plots and the second-highest with pie charts, followed by grouping scatter plots, bar charts, and stacked bar charts.
In contrast, for the few-shot prompt, the performance ranking, from highest to lowest accuracy, is observed in grouping scatter plots, pie charts, scatter plots, bar charts, and stacked bar charts.
In particular, the accuracy of the grouping scatter plots generation task exceeds 90\%.}

\begin{table}[h]
    \begin{center}
    \caption{
        \textbf{Performance of different prompt strategies:}
        The average performance of the three metrics of the entire Vega-Lite specifications output by evaluated LLMs using different prompt strategies on the nvBench.
    }
    \renewcommand\arraystretch{1.5}
    \begin{tabular}{cccc}
    \Xhline{1.2px} & \textbf{Vis Accuracy} & \textbf{Valid rate} & \textbf{Legal rate} \\
    \hline Few-shot & 95.49\% & 46.41\% & 25.83\% \\
    \hline Zero-shot & 96.02\% & 12.86\% & 6.23\% \\
    \hline Zero-shot-CoT & 96.21\% & 12.09\% & 6.90\% \\
    \hline PS-Plus-CoT & 96.11\% & 8.01\% & 4.56\% \\
    \hline Auto-CoT & 95.71\% & 14.83\% & 8.37\% \\
    \hline Least-to-Most & 95.32\% & 35.91\% & 19.10\% \\
    \hline Self-Refine & 95.34\% & 38.39\% & 20.20\% \\
    \hline Self-Consistency & 96.33\% & 65.75\% & 37.06\% \\
    \Xhline{1.2px}
    \end{tabular}
    \label{tab:total-results}
    \end{center}
\end{table}

\begin{table*}
\centering
\caption{
    \textbf{Total legality result:}
    The legality performance of evaluated prompt strategies with different LLMs or different chart types.
}
\captionsetup{width=0.8\textwidth}
    \centering
    \renewcommand\arraystretch{1.5}
    \begin{tabular}{@{}cccccccccc@{}}
    \Xhline{1.2px} & \textbf{Few-shot} & \textbf{Zero-shot} & \textbf{Zero-shot-CoT} & \textbf{PS-Plus-CoT} & \textbf{Auto-CoT} & \textbf{Least-to-Most} & \textbf{Self-Refine} & \textbf{Self-Consistency} & \textbf{AVG}\\
    \hline QWEN2 & 41.44\% & 6.88\% & 14.18\% & 12.19\% & 25.27\% & 21.61\% & 37.54\% & 50.87\% & 26.25\% \\
    \hline Llama3 & 36.18\% & 10.36\% & 15.92\% & 6.75\% & 22.93\% & 32.87\% & 16.56\% & 47.43\% & 23.63\% \\
    \hline GLM V4 & 27.13\% & 20.13\% & 1.44\% & 0.13\% & 0.08\% & 18.68\% & 21.10\% & 40.68\% & 16.17\% \\
    \hline Mistral & 23.91\% & 0.00\% & 9.85\% & 8.28\% & 1.83\% & 11.25\% & 21.83\% & 35.92\% & 14.11\% \\
    \hline Deepseek-R1 & 22.85\% & 0.00\% & 0.00\% & 0.00\% & 0.00\% & 21.61\% & 18.73\% & 33.97\% & 12.15\% \\
    \hline Gemma & 3.48\% & 0.00\% & 0.00\% & 0.00\% & 0.13\% & 8.58\% & 5.44\% & 13.46\% & 3.89\% \\
    \Xhline{0.8px}
    \hline Bar & 28.53\% & 7.38\% & 7.87\% & 5.27\% & 10.06\% & 22.93\% & 23.55\% & 41.39\% & 18.37\% \\
    \hline S-Bar & 7.85\% & 0.28\% & 1.93\% & 0.69\% & 0.55\% & 3.86\% & 4.55\% & 10.33\% & 3.75\% \\
    \hline Scatter & 21.99\% & 4.40\% & 11.11\% & 6.25\% & 6.71\% & 12.97\% & 12.73\% & 28.70\% & 13.11\% \\
    \hline G-Scatter & 17.62\% & 6.67\% & 10.48\% & 8.57\% & 8.10\% & 15.24\% & 7.14\% & 39.05\% & 14.11\% \\
    \hline Pie & 21.60\% & 0.00\% & 0.18\% & 0.00\% & 0.62\% & 0.09\% & 10.14\% & 24.52\% & 7.14\% \\
    \hline Line & 6.36\% & 4.39\% & 2.19\% & 1.10\% & 0.44\% & 4.39\% & 1.10\% & 11.62\% & 3.95\% \\
    \Xhline{0.8px}
    \hline Easy & 25.93\% & 5.51\% & 9.75\% & 5.99\% & 9.51\% & 15.80\% & 17.64\% & 37.49\% & 15.95\% \\
    \hline Medium & 33.24\% & 8.40\% & 7.46\% & 5.17\% & 11.06\% & 29.89\% & 27.76\% & 48.13\% & 21.39\% \\
    \hline Hard & 17.49\% & 4.93\% & 2.51\% & 1.89\% & 2.37\% & 4.98\% & 14.40\% & 23.19\% & 8.97\% \\
    \hline Extra Hard & 0.95\% & 0.00\% & 0.17\% & 0.26\% & 0.17\% & 0.60\% & 0.95\% & 1.38\% & 0.56\% \\
    \Xhline{0.8px}
    \hline \textbf{AVG} & 25.83\% & 6.23\% & 6.90\% & 4.56\% & 8.37\% & 19.10\% & 20.20\% & 37.06\% & 16.03\% \\
    \Xhline{1.2px}
    \end{tabular}
    \label{tab:total-legality}
\end{table*}

\subsubsection{Performance of different chart types and task difficulty}
\begin{figure}
    \centering
    \includegraphics[width=\columnwidth]{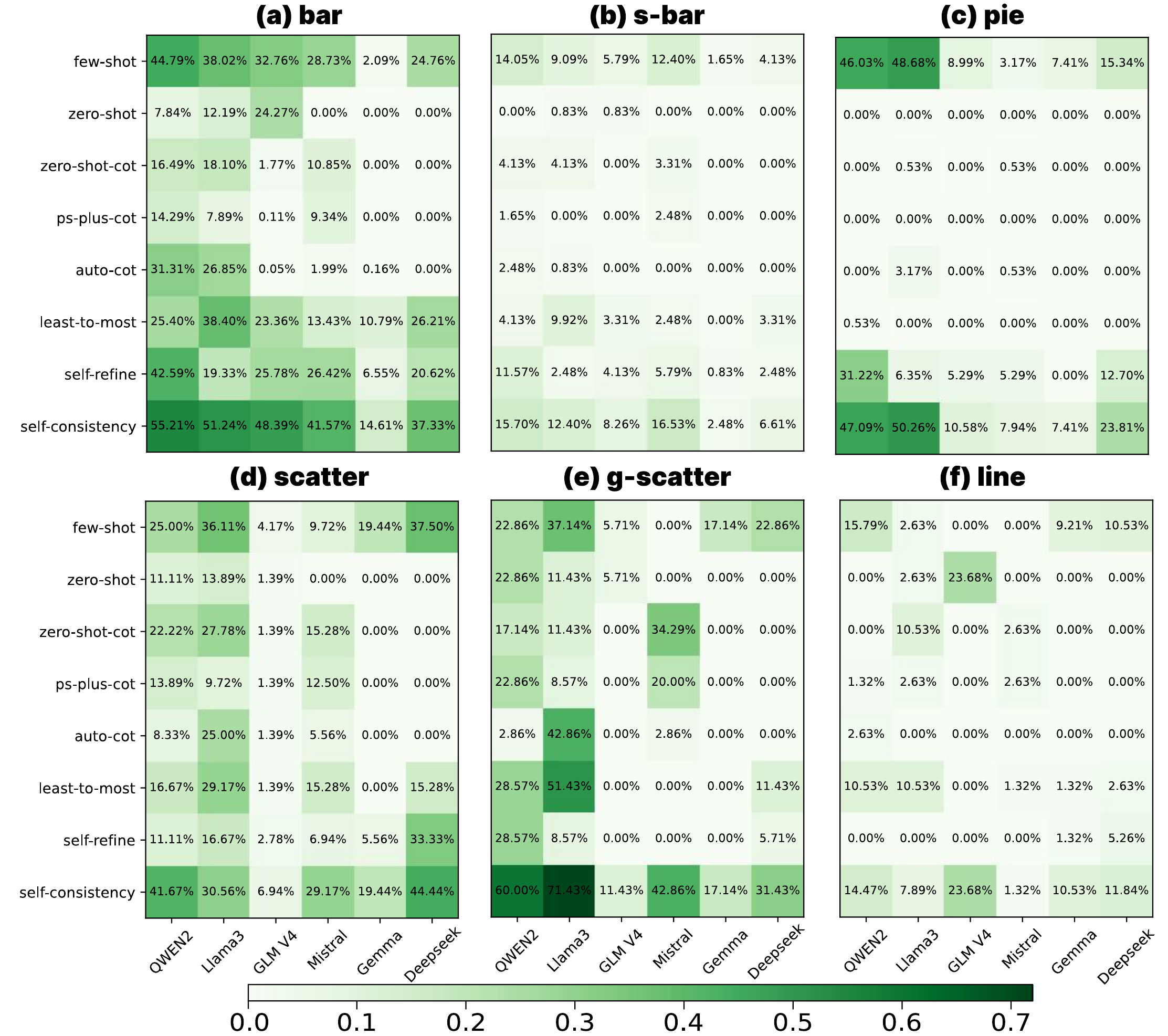}
    \caption{
        \textbf{Legality heat map of chart types}
        The figure presents the legality performance heat map of different chart types in the nvBench dataset.
    }
    \label{fig:type-heat-map}
\end{figure}

\revise{
Table~\ref{tab:total-legality} and Fig.~\ref{fig:type-heat-map} indicate that both chart types and task difficulty exhibit clear and internally consistent performance patterns across LLMs and prompt strategies.
When examined from the chart-type perspective, substantial variation is observed in legality performance. The approximate ordering is: bar charts\textgreater grouping scatter charts\textgreater scatter charts\textgreater pie charts\textgreater line charts and stacked bar charts.

Considering the interaction between prompt strategies and chart types, specific combinations can achieve comparatively stronger performance for particular visualizations while still adhering to the global type-based ranking. For example, Zero-shot-CoT yields the best performance for scatter charts, whereas PS-Plus-CoT is most effective for grouping scatter charts. For most other strategies, bar charts consistently produce the highest legality scores. 

From the perspective of task difficulty, the performance ranking is: medium tasks\textgreater easy tasks\textgreater hard tasks\textgreater extra hard tasks, reflecting the expected decline in performance with increasing complexity. However, an unexpected inversion appears between the easy and medium categories, with medium tasks outperforming easy tasks. This deviation from the anticipated hierarchy is attributable to the uneven distribution of pie chart, which is one of the lowest-performing chart type (see Fig.~\ref{fig:type-heat-map}c), across difficulty levels. 
As shown in Table~\ref{tab:total-legality}, pie charts constitute 30.31\% of the easy set but only 6.95\% of the medium set. 
This imbalance disproportionately suppresses the average legality of easy tasks while leaving the medium category relatively unaffected, thereby reversing the nominal difficulty-based ordering.
}

\subsubsection{Performance of different LLM models}
\begin{figure}
    \centering
    \includegraphics[width=\columnwidth]{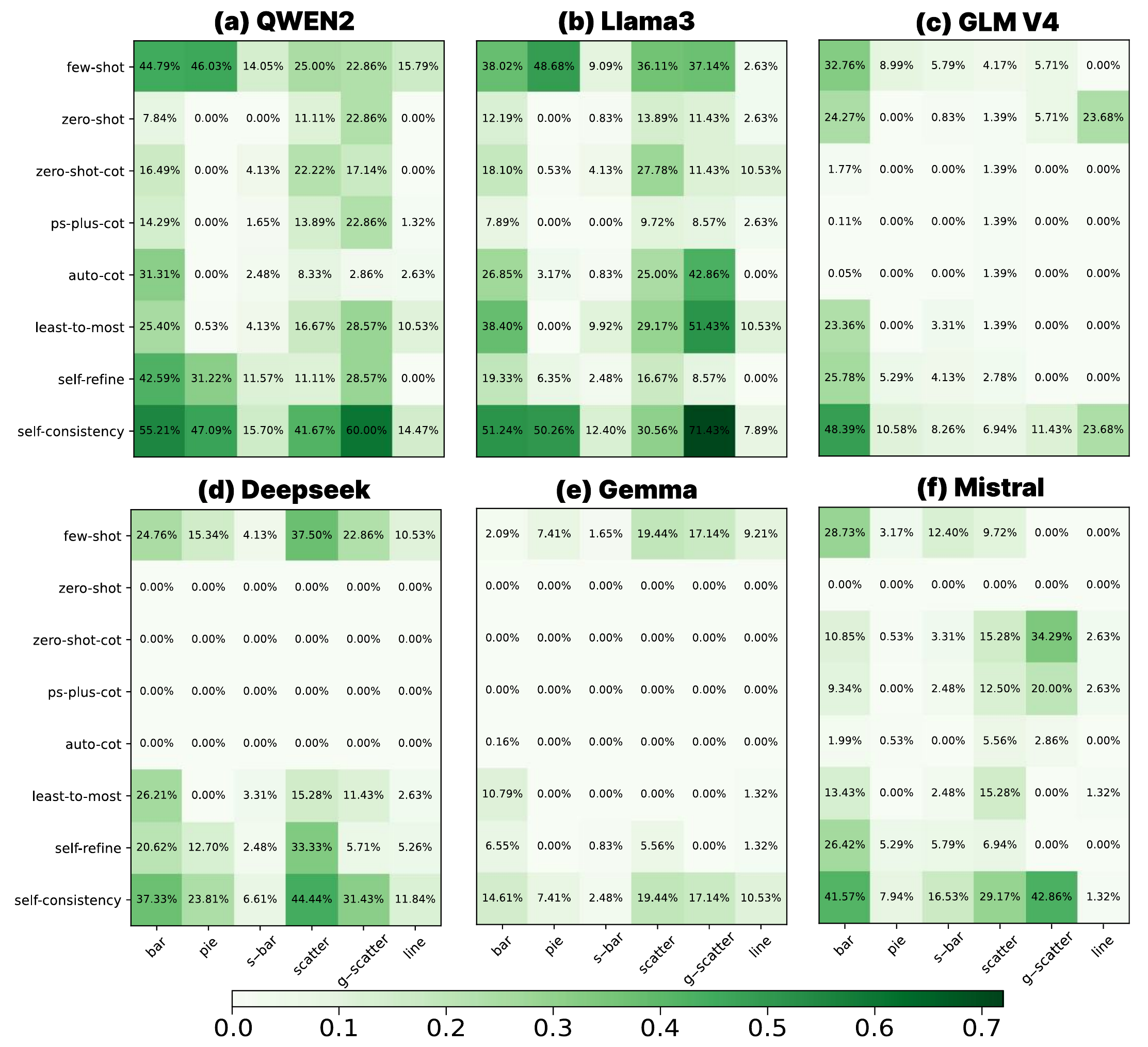}
    \caption{
        \textbf{Legality heat map of LLMs.}
        The figure presents the legality performance heat map of different LLMs.
    }
    \label{fig:model-heat-map}
\end{figure}

\revise{
Different LLMs exhibit marked disparities in legality performance on the NL2VIS task. As shown in Table~\ref{tab:total-legality} and Fig.~\ref{fig:model-heat-map}, the approximate performance ranking is: Qwen2, Llama3\textgreater GLM V4, Mistral, DeepSeek-R1-Llama3.1\textgreater Gemma. Although such differences are evident, their underlying causes remain unclear because the detailed architectures, pre-training corpora, and alignment procedures of these proprietary models are not fully disclosed. One notable anomaly is the weak performance of the Self-Refine strategy on Llama3, which we attribute to the model’s limited ability to interpret the iterative refinement structure embedded in the Self-Refine prompt; this observation is examined in greater detail in Section~\ref{subsection:deep-findings}.

Further, our evaluation shows that Qwen2, Llama3, and GLM V4 consistently outperform Mistral, Gemma, and DeepSeek-R1-Llama3.1 across most prompting conditions. 
In particular, for all strategies except Few-shot, the latter three models often produce legality scores close to zero, indicating difficulty in handling structured NL2VIS instructions without explicit contextual examples.

Although Fig~\ref{fig:prompt-heat-map} suggest that prompt strategies do not alter the overall performance hierarchy across model–chart-type combinations as we discussed in Section~\ref{sssec:prompt}, they still interact with model characteristics in nuanced and model-dependent ways.
Reasoning-oriented prompts such as Zero-shot-CoT and PS-Plus-CoT tend to enhance the performance of models with stronger multi-step reasoning capabilities, whereas models such as Gemma with weaker instruction-following abilities show minimal gains under the same conditions. 
Considering models and prompt strategies jointly, several combinations demonstrate notably strong performance: Zero-shot-CoT performs particularly well with GLM V4, whereas Least-to-Most yields the strongest results on Llama3. 
Across the broader set of model–prompt combinations, Qwen2 generally achieves the highest legality scores.

Collectively, these results indicate that baseline performance is primarily shaped by model architecture and pre-training, yet the alignment between model reasoning capabilities and prompt design can further modulate performance and produce substantial gains in specific scenarios.
}

\section{Findings}

In this section, we first 
\delete{categorize the factors that constrain the performance of GPT models on NL2VIS tasks}
\revise{investigate several intriguing patterns identified in the evaluation results. 
By systematically analyzing their underlying mechanisms, we summarize key insights that may guide future research on the NL2VIS task.} 
Subsequently, we delineate the limitations identified within the existing NL2VIS benchmark.
\begin{figure*}
    \centering
    \includegraphics[width=\textwidth]{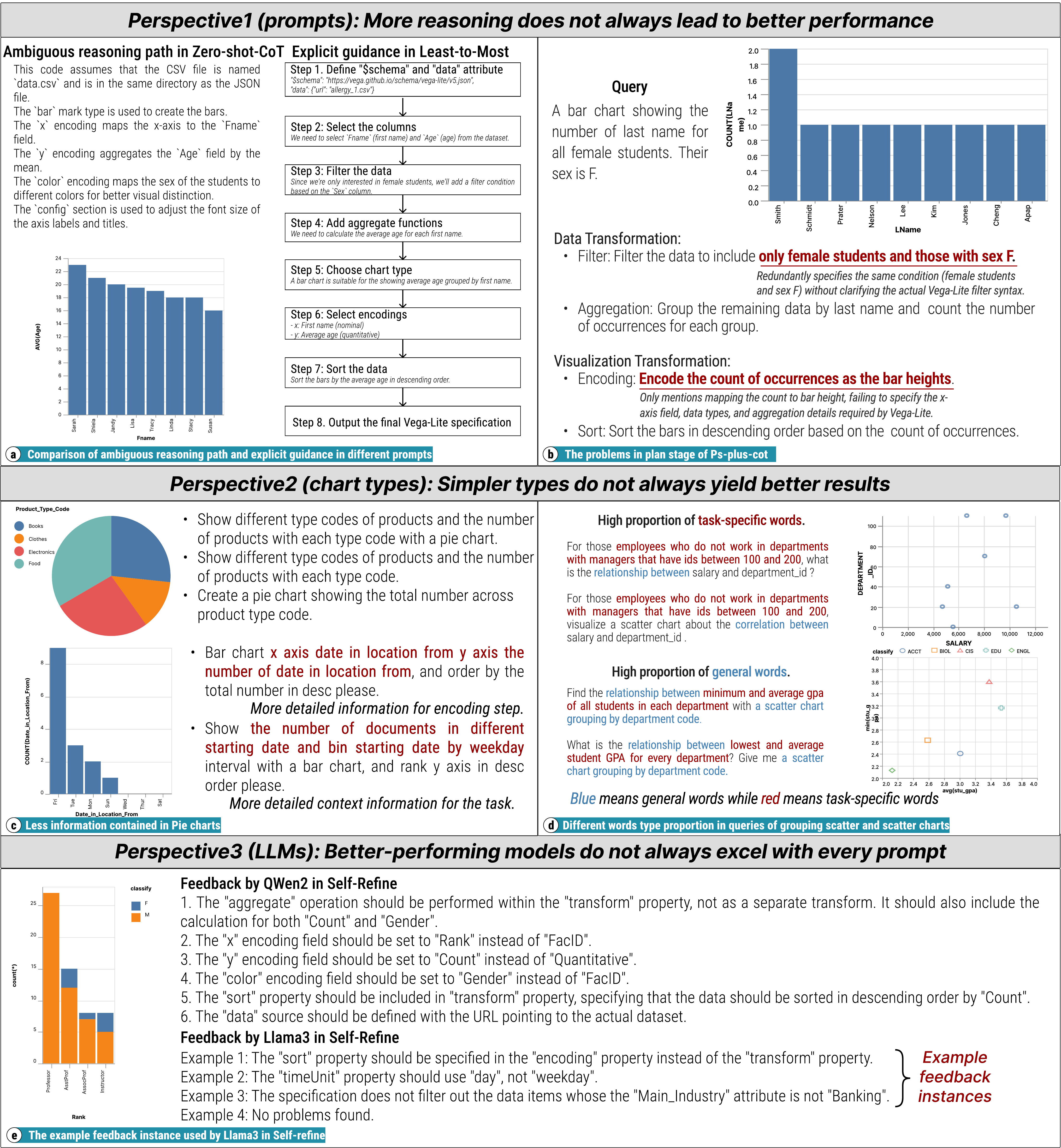}
    \caption{
        \revise{\textbf{Counterintuitive findings observed in results.} 
        The figure presents illustrative examples corresponding to the findings discussed in Section~\ref{subsection:deep-findings}. 
        First, regarding prompting strategies: 
        (a) illustrates ambiguous reasoning trajectories that emerge under the Zero-shot-CoT and Least-to-Most prompting strategies. 
        (b) highlights issues in the plans generated by LLMs. 
        Second, regarding chart types: 
        (c) shows comparisons of queries from pie charts and bar charts, where bar-chart queries contain richer and more fine-grained information than pie-chart queries. 
        (d) depicts queries for scatter charts and grouped scatter charts, showing that scatter-chart queries exhibit a high proportion of task-specific terminology, whereas grouped-scatter queries contain a higher proportion of general-purpose words. 
        Third, regarding LLM behavior: 
        (e) presents the first three feedback examples used by Llama-3 within the Self-Refine framework.}
    }
    \label{fig:deep-findings}
\end{figure*}
\subsection{\revise{Counterintuitive Findings Observed in Evaluation Results}}
\label{subsection:deep-findings}
\delete{ The previous section shows the performance of GPT-3.5 and GPT-4 models on the nvBench dataset. 
We analyze errors to find the limitations of GPT models on the Vega-Lite generation task, and then divide them into the following three categories to facilitate further optimization.}

The previous section presents the performance of \revise{LLMs under different prompting strategies} on the nvBench dataset. Building on these results, we further analyze \revise{the core patterns that emerge in the NL2VIS task. Although the overall evaluation framework in Section~\ref{subsection:results} includes four dimensions, the task-difficulty labels in nvBench are inherently subjective and thus less suitable for in-depth interpretation. 
Consequently, our findings focus on the three more objective and informative dimensions, which we distill into key perspectives to guide subsequent optimization.}

\revise{\textbf{Perspective 1: In terms of prompt, more reasoning does not always lead to better performance.}
From the perspective of prompt design, our experiment employed eight distinct prompting strategies. 
Among the strategies tested, few-shot prompting stood out as particularly effective, outperforming all other approaches except Self-Consistency. 
Its advantage suggests that providing explicit, high-quality Vega-Lite exemplars improves not only the model’s task reasoning but also its grasp of the underlying Vega-Lite grammar. 
To further illustrate the influence of prompt structure, we examine the behavior of Least-to-Most and PS-Plus-CoT prompts, which reveal how different forms of reasoning guidance shape model outputs.
These findings underscore the need for a more nuanced approach to prompt design, offering insights for future research in task-specific prompting strategies.}

\delete{(c) Inadequate understanding of Vega-Lite transformation parameters.
The type of error involves misusing the transformation operations within the Vega-Lite specification.
For example, as depicted in Fig.~\ref{fig:errors-LLMs}(c), the generated specification may have a correct JSON format and not use non-existent operations. However, the use of the \vlbox{bin} in the \vlbox{transform} property is incorrect, resulting in an output specification that violates the Vega-Lite grammar.
In this example, the task query requests binning \textit{date\_address\_to} attribute by weekday, achievable by defining \vlbox{timeUnit} in the \vlbox{transform} property.
The output provided by GPT models confuses the function of \vlbox{bin} and \vlbox{timeUnit}, resulting in an invalid specification. }

\delete{The inaccuracies observed in Vega-Lite specifications highlight the non-negligible issue of hallucination within LLMs, as outlined in previous research~\cite{OpenAI-arxiv230308774-GPT4report}.
Hallucination problems manifest in the Vega-Lite output, wherein LLMs may specify data transformations not explicitly mentioned in queries or introduce grammar errors.
The examples presented underscore the presence of grammar errors within Vega-Lite visualization specifications, suggesting the requirements for an effective linting strategy to enhance performance.
Linting techniques tailored for the output of LLMs should prioritize scrutinizing the transformations within the Vega-Lite specification.}

\revise{\textit{(a) LLMs lack the knowledge of Vega-Lite grammar.}}
\revise{Evaluation results reveal that the Few-shot prompt strategy consistently outperforms all other prompt methods except Self-Consistency.
Due to the substantial computational overhead associated with the Self-Consistency strategy, we conclude that Few-shot prompt offers the most balanced trade-off between correctness and computational efficiency.
Unlike reasoning-oriented strategies, such as PS-Plus-CoT, Least-to-Most, Zero-shot-CoT, and Auto-CoT, which we categorize as reasoning-oriented because they explicitly guide the model through stepwise or hierarchical decision-making processes, Few-shot prompting primarily compensates for the model’s insufficient knowledge of Vega-Lite grammar and conventions. 
This distinction highlights two fundamentally different intervention levels: while reasoning-oriented prompts improve how the model reasons about visualization design, Few-shot prompt enhances what the model knows about the syntactic of Vega-Lite declarative grammar.

The evaluation outcomes further suggest that the main bottleneck in the NL2VIS task lies not in the model’s reasoning capability, but in its limited understanding of Vega-Lite grammar.
Because LLMs are predominantly trained on natural language and procedural programming languages such as Python or Java, their exposure to specialized declarative grammars like Vega-Lite remains limited.
As a result, reasoning-oriented prompts fail to bridge this syntactic representational gap, whereas explicit exemplars in few-shot settings directly ground the model’s generation in correct syntactic and semantic patterns.

This finding carries an important implication for future research: when applying LLMs to NL2VIS tasks, it is more effective to address the model’s representational understanding of domain-specific languages rather than merely refining its reasoning pathways.
Hence, prompt designs should prioritize direct exemplification over abstract reasoning cues, particularly for formal visualization grammars like Vega-Lite, where syntactic grounding exerts a stronger influence on model accuracy and stability than stepwise reasoning.}

\revise{\textit{(b) Least-to-Most outperforms CoT series in multi-stage visualization tasks.}
Evaluation results consistently demonstrate that across all evaluated LLMs, the Least-to-Most prompting strategy outperforms the CoT and its variants as we classified in Section~\ref{sec:prompt-strategies} in the NL2VIS task. 
This advantage arises from the alignment of the Least-to-Most prompt with tasks that are inherently sequential or multi-stage.~\cite{zhouLeasttoMostPromptingEnables2023}. 
Unlike CoT and its variants, which encourages implicit and ambiguous reasoning paths, Least-to-Most provides explicit, stage-wise guidance as it is shown in Fig.~\ref{fig:deep-findings}a, progressively breaking down a complex goal into smaller, logically ordered sub-tasks. 
Such structured prompting enhances both the interpretability and stability of model behavior during visualization code generation.

This observation aligns with the conceptual framework proposed by LIDA, which interprets NL2VIS as a four-stage process encompassing comprehension, mapping, generation, and refinement~\cite{Dibia-acl2023-LIDA}. 
The Least-to-Most strategy reflects this structure by incorporating human-like hierarchical reasoning into the generation process, ensuring that each stage builds coherently upon the previous one. 
In contrast, CoT and its variants often lack clear procedural segmentation, leading to redundant explanations or inconsistent intermediate outputs.

This finding suggests that prompts incorporating explicit, human-designed task decomposition hold significant promise for NL2VIS tasks based on LLMs. The superior performance of the Least-to-Most prompt strategy implies that future research should further explore prompting paradigms that integrate domain knowledge or process-aware scaffolding into instruction design.}

\delete{(d) Inadequate understanding of Vega-Lite operation parameters.
This error type indicates that LLMs fail to fully comprehend the operation parameters of Vega-Lite.
Unlike the specifications outlined in Limitation 1, this type exhibits no grammar errors but still produces an incorrect visualization compared to the ground truth.
For example, a discrepancy arises from the incomplete understanding of the \vlbox{timeUnit} parameters used in the \vlbox{transform} property.
Although the query mandates binning \textit{date\_address\_to} by weekday, GPT models incorrectly assign the \vlbox{timeUnit} parameter value as \textit{weekday} instead of the correct parameter \textit{day}.
This instance highlights the necessity for LLMs to possess comprehensive knowledge of the Vega-Lite grammar and underscores their susceptibility to utilizing incorrect parameter assignments influenced by the query.}

\delete{This type of error implies that one aspect of improving the performance of LLMs on the， \revise{NL2VIS} task is to develop more strategies to facilitate LLMs to understand and correctly use the parameters of Vega-Lite specifications.
For example, the documentation of Vega-Lite~\cite{vega-lite-document} should be one important resource for realizing this target.}

\revise{\textit{(c) The plan stage hurts PS-Plus-CoT performance.}
Evaluation results show that the PS-Plus-CoT prompting strategy consistently underperforms the Zero-shot-CoT approach across all evaluated LLMs as it is shown in Table~\ref{tab:total-legality}. 
This finding contradicts the conclusions drawn in the original PS-Plus-CoT study~\cite{wangPlanandSolvePromptingImproving2023}. 
However, given the significant differences between the dataset used in that study and the NL2VIS dataset, especially in terms of task formulation, this inconsistency is not entirely unexpected.

Statistical analysis further reveals that the predominant error type is the Vega-Lite format error, indicating that grammatical fidelity is compromised by the inclusion of an additional planning stage.
Prior studies~\cite{geng2025control} suggests that when a specific stage in a multi-step reasoning process cannot provide sufficient meaningful contribution to the target task, it can inadvertently disrupt the model’s attentional coherence, leading to degraded performance in subsequent stages. 
As shown in Fig.~\ref{fig:deep-findings}b, the planning stage in PS-Plus-CoT often faces issues such as redundant or ambiguous condition declarations in the Filter step, and misaligned objectives in the \vlbox{encoding} step, which focus on explaining visual semantics rather than specifying encoding mappings. 
Consistent with this theoretical perspective, we argue that the plan stage in PS-Plus-CoT provides limited or even counterproductive guidance for generating Vega-Lite specifications in the NL2VIS task. 
As a result, the model’s attentional focus on syntactic constraints weakens, resulting in decreased Vega-Lite format accuracy.}

\revise{In summary, our findings emphasize that the quality and relevance of guidance are more critical than its complexity. 
Among various prompting strategies, direct exemplification such as Few-shot proves to be the most effective approach for NL2VIS, emphasizing the need for concise, contextually grounded guidance in Vega-Lite specification generation. 
Moreover, structured stage-wise reasoning such as Least-to-Most provides meaningful guidance, whereas ambiguous and implicit plans such as PS-Plus-CoT may hinder performance. }

\revise{\textbf{Perspective 2: In terms of chart types, easier types do not always yield better results.}
Intuitively, LLMs are expected to achieve higher visualization accuracy when generating Vega-Lite specification for simpler charts. However, evaluation results reveal that the visualization accuracy (vis accuracy) of LLMs is not necessarily the highest for such simplified visualizations. This discrepancy is particularly evident in the case of pie charts under Zero-shot prompt, as well as when comparing scatter charts to grouping scatter charts.}

\delete{\textit{(a) JSON format error.} 
This error category involves misusing the properties of the Vega-Lite specification, resulting in an invalid JSON object.
As depicted in Fig.~\ref{fig:errors-LLMs}(a), examples generated by GPT models may violate the format rules of valid JSON objects.
For example, GPT models may incorrectly place the properties \vlbox{mark} and \vlbox{encoding} within the transform property, erroneously treating the property \vlbox{encoding} as the end of the specification.
This misuse often leads to a missing square bracket at the end of the \vlbox{transform}, causing a violation of the JSON format requirements.}

\begin{table}[h]  %
    \begin{center}
            \caption{
        \textbf{Average query length:}
        The average word length of the four chart types queries on the nvBench.
    }
    \renewcommand\arraystretch{1.5}
    \begin{tabular}{cccc}  %
    \Xhline{1.2px} 
    \textbf{Pie} & \textbf{Line} & \textbf{Bar} & \textbf{Stacked Bar} \\
    \hline
    15.65 & 25.42 & 25.67 & 28.33 \\
    \Xhline{1.2px}
    \end{tabular}
    \label{tab:query-state}
    \end{center}
\end{table}

\revise{\textit{(d) Poor performance of pie charts under Zero-shot prompt and its variations.} 
Evaluation results reveal that pie charts consistently perform weakly across Zero-shot, Zero-shot-CoT, PS-Plus-CoT, Auto-CoT, and Least-to-Most, as shown in Fig.~\ref{fig:type-heat-map}(c). 
Interestingly, the average length of Vega-Lite specification for pie charts is relatively short, which contradicts intuitive expectations that simpler charts should yield better results.

To investigate this counterintuitive finding, we analyzed the nvBench dataset and computed the average query length for different chart types. The results shown in Table~\ref{tab:query-state} indicate that queries associated with pie charts are generally shorter than those of other chart types. 
Under conditions without exemplar-based prompting, the query serves as the only source of task understanding for LLMs; 
thus, longer queries tend to provide richer contextual information as it is shown in Fig.~\ref{fig:deep-findings}c, enabling the LLM to better comprehend the NL2VIS task and achieve higher accuracy.

Furthermore, the error analysis result indicates that 77.69\% of the incorrect outputs are due to Vega-Lite format errors. 
Upon closer inspection, we found that most of these errors arise from the model repeatedly using $x$ and $y$ \vlbox{encodings} inappropriately. 
Since other chart types in our experiments, such as bar, line, and scatter charts, are typically encoded along $x$–$y$ axis, it is likely that the model’s pretraining data contained fewer examples of charts without $x$–$y$ \vlbox{encodings} (e.g., pie charts). 
Consequently, the model tends to overgeneralize by applying $x$–$y$ \vlbox{encodings} to chart types that do not use them.

The above observation suggests that, when employing LLMs to generate Vega-Lite specifications for charts without $x$–$y$ \vlbox{encodings} (such as pie charts or treemaps), providing an explicit example in the prompt can effectively guide the model toward producing more accurate and task-aligned code.}

\delete{\textit{(b) Using non-existent properties in Vega-Lite specifications.}
This error type involves adding nonexistent properties within the generated specification.
Fig.~\ref{fig:errors-LLMs}(b) illustrates an example that conforms to JSON format requirements but utilizes a nonexistent \vlbox{sort} property in the \vlbox{transform} property.
According to the task description, the chart should sort the \textit{Director} attribute in ascending order, which can be realized by defining the \vlbox{sort} in the \vlbox{x} property as highlighted in bold text in Fig.~\ref{fig:errors-LLMs}(b). 
Despite GPT models correctly understanding the task query to define the sorting operation, the misuse of the \vlbox{sort} property in the \vlbox{transform} property leads to an incorrect specification.
Moreover, errors with respect to the \vlbox{sort} property frequently occur in tasks that require the sorting of one attribute, contributing to the relatively low performance of GPT models in such tasks.}

\revise{\textit{(e) Better performance of grouping scatter charts comparing with scatter charts.}
Evaluation results demonstrate that grouping scatter charts (see Fig.~\ref{fig:type-heat-map}(e)) generally outperform regular scatter charts (see Fig.~\ref{fig:type-heat-map}(d)), a trend that is particularly evident among LLMs with better results. 
According to statistics from the nvBench dataset, the average query length of grouping scatter is comparable to that of scatter, and the average Vega-Lite specification length for grouping scatter is longer, indicating a higher level of grammatical complexity. 
Despite this increased complexity, grouping scatter still achieves better performance, suggesting that the performance gap is not driven by specification length or task complexity, but rather by differences in semantic characteristics.

To further explore this, we conducted a word frequency analysis of the queries associated with both chart types. 
The results show that grouping scatter queries more frequently contain general terms such as ``group'', ``attribute'', and ``correlation'', which commonly appear across multiple tasks. 
In contrast, the top three frequent terms in scatter queries are ``correlation'', ``employees'', and ``id'', with the latter two being task-specific. 
To provide a clearer understanding of this observation, Fig.~\ref{fig:deep-findings}(d) presents representative examples of scatter and grouping scatter queries. 
Based on the above findings, we speculate LLMs tend to achieve higher accuracy when dealing with prompts containing shared and semantically general vocabulary, as these facilitate more consistent internal task representations.

These findings suggest that, in future NL2VIS tasks using LLMs, prompt design should prioritize alignment with the overall semantic structure of the task, rather than focusing narrowly on individual instances. 
By providing prompts that are representative and semantically broad, we can better guide the model's understanding of the task's intent, ultimately enhancing both the robustness and accuracy of the generated Vega-Lite specifications.}
\revise{Overall, the low accuracy of pie charts highlights the challenges in handling grammatically distinct Vega-Lite specifications compared to the examples provided in the prompt, while the superior results of grouping scatter plots indicate strength in tasks with semantically shared elements. 
Future work should focus on aligning prompts with task semantics, using representative examples to guide models toward generating more accurate and robust Vega-Lite specifications.}

\delete{Limitation 2: Inadequate understanding of the Vega-Lite parameters.
This category indicates that the Vega-Lite specifications generated are devoid of grammar errors and can be successfully transformed into visualizations. 
However, this limitation leads to the ``Chart Content Mismatch'' error as documented in Table~\ref{tab:errors}.
Despite the successful transformation into visualizations, discrepancies arise between the generated visualizations and the ground truth due to the GPT model's limited comprehension of the parameters within Vega-Lite.}

\revise{\textbf{Perspective 3: In terms of models, better-performing models do not always excel with every prompt.}}
\revise{
Interestingly, although Llama3 generally ranks among the top-performing models and the Self-Refine strategy is considered highly effective, the accuracy achieved under this combination is relatively lower than expected. 
This relative underperformance is treated as an anomalous observation: one would anticipate a top-tier model paired with an effective prompt to produce correspondingly high results. 
The finding highlights that even strong models may exhibit suboptimal performance when prompt workflows are misinterpreted, particularly in tasks requiring iterative or multi-step reasoning. Achieving optimal NL2VIS outcomes thus depends not only on the inherent strength of the model or prompt, but critically on their precise alignment and compatibility.

A closer inspection reveals the underlying cause of this anomaly. As described in Section~\ref{sec:evaluated-prompts}, our Self-Refine prompt template includes three example feedback instances, after which the model is expected to iteratively improve visualization generation based on feedback. Fig.~\ref{fig:deep-findings}e shows that instead of generating a new feedback instance based on the three provided examples, Llama3 produced its own feedback and mistakenly treated it along with the three given examples as four tasks to be completed in the subsequent refinement step. This misunderstanding caused the model to loop over redundant reasoning paths rather than improving the visualization output.

Furthermore, this phenomenon underscores broader considerations for NL2VIS method design. High-performing models cannot be assumed to generalize effectively across all prompt types, especially those requiring complex reasoning or feedback-based refinement. Selecting models for NL2VIS tasks should therefore consider not only benchmarked accuracy or model scale but also the compatibility between the model’s reasoning characteristics and the cognitive demands imposed by the prompt. For example, tasks involving iterative refinement or multi-step decision making demand models with strong instruction-following and contextual comprehension capabilities. These insights emphasize that successful NL2VIS generation depends on the synergy between model capabilities and prompt design, rather than solely on individual performance metrics.

Overall, the case of Llama3 under Self-Refine serves as a cautionary example: optimizing NL2VIS systems requires attention to nuanced model-prompt interactions, and high baseline model performance does not automatically translate to success across all prompting scenarios.}

\subsection{Reexamining NL2VIS Benchmark}

We scrutinize the discrepancies in the \delete{GPT model} \revise{LLMs} outputs and discover that certain instances within the nvBench dataset contain inaccuracies, attributing some of the \delete{GPT model} \revise{LLMs} failures to shortcomings within the nvBench. 
\delete{We classify the reasons for these failures into three categories: incorrect query statements, improper data mapping, and incorrect underlying data/labels.}
We analyze the failures and introduce insights to refine the benchmark design for the NL2VIS task. 

\begin{figure*}
    \centering
    \includegraphics[width=\textwidth]{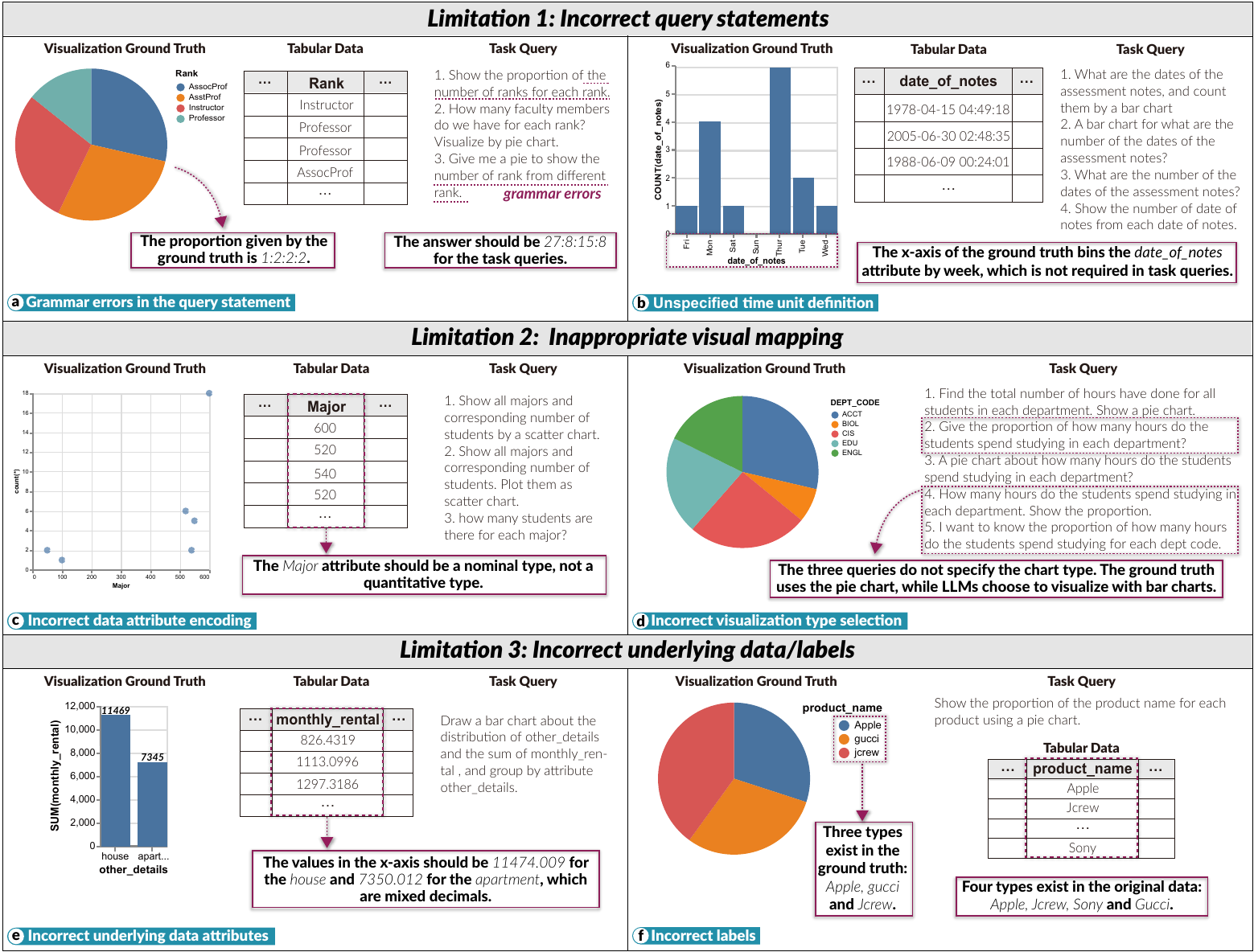}
    \caption{
        \textbf{Limitations of the nvBench dataset.} 
        The first limitation involves incorrect query statements, encompassing  
        (a) grammar errors in the query statement and (b) Unspecified time unit definition in the query statement;
        The second limitation is inappropriate data mapping, including 
        (c) incorrect data attribute mapping and (d) incorrect visualization type selection; 
        The third limitation is incorrect underlying data and labels, comprising 
        (e) incorrect underlying data attributes and (f) incorrect labels;
        The purple dashed box indicates the location of the mistakes, with explanations provided in the purple solid box.
    }
    \label{fig:limitaion-benchmark}
\end{figure*}

\textbf{Limitation 1: Incorrect query statements.} 
This type of limitation indicates that queries cannot match the ground truth visualizations. 
\delete{We summarize these mistakes as ``incorrect query statements'', because the query statements of these instances either contain some grammar errors or can be easily modified to match the provided ground truth visualization.}

(a) \textit{Grammar errors in the query statement.}
A notable issue in the benchmark is that some query statements contain grammatical errors or ambiguous phrasing, which leads to inconsistencies between their intended meaning and the corresponding ground truth visualizations. These inaccuracies introduce confusion and reduce the reliability of the evaluation.
The first example, shown in Fig.~\ref{fig:limitaion-benchmark}(a), illustrates such a mismatch. The query \textit{``Show the proportion of the number of ranks for each rank''} ostensibly asks for the number of faculty members in each rank category, for which the expected values would be 27 (Professors), 8 (Associate Professors), 15 (Assistant Professors), and 8 (Instructors). However, the ground truth instead reports a 1:2:2:2 distribution, which actually denotes gender proportions for each rank. Thus, the query is misleading: the correct statement should request the proportion of genders rather than the count of ranks.
\delete{Remarkably, GPT models \revise{LLMs} successfully provide the correct answer of 27:8:15:8.}
\delete{A plausible explanation is that the ground truth chart accurately captures the underlying intent of this instance, while the queries themselves are erroneous.
The query aligning with the ratio 1:2:2:2 pertains to the distribution of genders for each rank. 
Specifically, the ``professors'' category features only one gender, whereas the other three categories encompass two genders. 
Consequently, the correct queries should inquire about the proportion of genders for each rank.}

(b) \textit{Unspecified time unit definition in the query statement.}
Unspecified time unit definition poses a challenge in certain nvBench instances that requires temporal data presentation in visualization results. 
When queries do not explicitly define the necessary time unit (e.g., seconds, hours, days), discrepancies arise between the temporal units inferred by \delete{GPT models}\revise{LLMs} and those used in the ground truth, which consequently lowers accuracy.
For example, Fig.~\ref{fig:limitaion-benchmark}(c) illustrates an instance where the time unit is unspecified. 
The ground truth visualization bins the \textit{date\_of\_notes} attribute, which is measured in seconds within the dataset, into discrete days of the week. 
However, this temporal requirement is neither explicitly nor implicitly mentioned in the query statements. \delete{for example, ``\textit{What are the number of the dates of the assessment notes?}'', leading GPT models \revise{LLMs} to generate a visualization result significantly divergent from the ground truth.}

Mismatched queries that fail to align with the visualizations pose a significant challenge in NL2VIS benchmarks. 
\delete{Whether the query is manually designed or automatically generated, completely bypassing this issue can cause serious problems, thus highlighting the necessity of reviewing benchmark datasets, like making the query statement clear and  free of syntax errors.}
Given that certain foundational models possess multi-modal capabilities, enabling them to comprehend both natural language and images simultaneously, leveraging such models can facilitate effective examination of query-visualization content alignment.

\textbf{Limitation 2: Inappropriate visual mapping.}  
\delete{Data mapping involves encoding data attributes into appropriate visual channels and selecting the correct chart types for visualization. }
\revise{Studies have consistently shown that} effective visualizations rely on suitable chart types and employing appropriate data mapping strategies~\cite{munzner2014visualization, apt1986tog, assessing2018cgf}. \delete{which have been extensively studied and summarized in the  community~\cite{munzner2014visualization, apt1986tog, assessing2018cgf}.}
However, our analysis of the nvBench dataset revealed two prevalent issues related to incorrect visual encoding and chart selection, which compromise the benchmark's validity.

(c) \textit{Incorrect data attribute encoding}. 
This type of mistakes occur when data attributes are encoded into incorrect visual channels due to being treated as the wrong data type.
For example, in a scatter plot, as shown in Fig.~\ref{fig:limitaion-benchmark}(c), the ground truth presents the number of different majors on the \textit{x}-axis, treating it as a quantitative type. 
However\delete{, since the number of different majors typically consists of discrete integer values rather than continuous numbers}, it should ideally be classified as a nominal type. 
Another scenario of data type errors involves the misclassification of temporal data as nominal type in the ground truth of nvBench. 
This issue is particularly common in nearly every instance involving temporal data in the benchmark. 
\delete{Such misclassification can lead to misleading data insights, as organizing nominal type data alphabetically may result in an inaccurate time sequence. 
Therefore, it’s essential to accurately classify temporal data to ensure precise representation and interpretation. }

(d) \textit{Incorrect visualization type selection}.
\delete{Most queries in the nvBench demand a specific chart type, but some do not state a chart type explicitly, which negatively impacts the judgment of LLMs GPT models.}
\revise{We find that some queries in the nvBench do not state a chart type explicitly.}
The instance shown in Fig.~\ref{fig:limitaion-benchmark}(b) has five queries, and only the first and third queries specify the pie chart requirement explicitly. 
\delete{GPT models}\revise{LLMs} generate pie charts for these two queries (the first and the third) and give out bar charts for the other three queries.
For these three queries that do not require pie charts explicitly, they emphasize the proportion and quantity in the statement simultaneously. 
Thus, bar charts are also appropriate.
\delete{We define the mismatch between generated results and ground truths as incorrect results.
Thus, the problem of unstated chart types also decreases the performance of GPT models on the nvBench.}
\revise{We conducted an in-depth analysis of the specific proportions and found that this ratio reached as high as 18.39\%, which could substantially affect the accuracy of LLMs in performing the NL2VIS task.}

The queries in the nvBench are generated automatically by an NL2SQL-to-NL2VIS synthesizer~\cite{Luo-sigmod2021-NL2VISbenchmark}, and we suppose the generation procedure for natural language queries can be further optimized to eliminate the ambiguity of chart type statements existing in a part of queries.
\delete{For instance, we can use LLMs to examine queries and add the most appropriate chart type, which is determined by the LLM depending on the rules of chart type selection we input in the prompt, to the problematic query.}
\delete{To address inappropriate data mappings in visualizations, we can utilize linting techniques similar to those used to rectify invalid Vega-Lite specifications, as explained in Sec.~\ref{subsection:questions-of-LLM}. }
By establishing rules regarding data mappings and employing linting techniques to validate specifications in the benchmark, we can identify and correct inappropriate data mappings.

\textbf{Limitation 3: Incorrect underlying data/labels.}
\delete{As discussed in Sec.~\ref{sec:metrics}, we evaluate the capabilities of GPT models by computing the matching accuracy between ground truths and generated visualizations. 
The visualizations encompass visual elements with underlying data attributes and labels for explanations.
This type of mistake implies that the mistakes lie not in the GPT-generated results but rather in the ground truth, where either the underlying data attributes or the labels for explanations are incorrect.}
\revise{As described in Section~\ref{sec:metrics}, we assess LLMs by measuring the matching accuracy between generated and ground-truth visualizations.
Such errors indicate flaws in the ground truth rather than in the LLMs outputs, caused by incorrect data attributes or explanatory labels.}

(e) \textit{Incorrect underlying data attributes.} 
Many instances in the nvBench dataset involve numerical calculations, such as summation and averaging, where the decimal part of the data in the original tables is directly omitted before calculation. 
This practice leads to different computation results between the ground truths and the output of \delete{GPT models}\revise{LLMs}.
As shown in Fig.~\ref{fig:limitaion-benchmark}(e), the \textit{monthly\_rental} attribute in the original data table contains the mixed decimal.
However, in the Vega-Lite specification provided by the nvBench, shown in the rightmost, the final summation results are the integer number.
Upon examination, we found that the nvBench drops the decimal part of the \textit{monthly\_rental} attribute before summation, resulting in an error greater than one compared with correct answers.

(f) \textit{Incorrect labels.} 
Another issue found in the nvBench ground truths is incorrect labels.
Fig.~\ref{fig:limitaion-benchmark}(d) exemplifies this mistake.
In this instance, the task requires determining the proportion of products for each product name. 
However, the ground truth chart notably omits the \textit{Sony} label from the original table.
In contrast, the output of \delete{GPT models} \revise{LLMs} is correct.
This discrepancy could be attributed to some mistakes during the generation of this instance, as the Vega-Lite specification in the ground truth adopts the inline data source.

To ensure consistency between the data displayed in the visualization and the original table, we need to ensure that the content of the \vlbox{data} property in Vega-Lite specifications matches that of the original table. 
Alternatively, we can opt to use the inline data source in Vega-Lite specifications.

\delete{While the analysis of Limitations in nvBench provides insights into the challenges and shortcomings of NL2VIS benchmarks, it is essential to recognize that these issues may not be unique to this specific dataset.
Other benchmark datasets in the NL2VIS domain may also suffer from similar Limitations, albeit with variations in severity and frequency. 
Therefore, the above Limitation analysis can provide the foundation for systematically analyzing and identifying potential Limitations in other benchmark datasets.
Furthermore, it can help to establish a comprehensive understanding of the limitations and areas for improvement in the evaluation of NL2VIS models. 
By conducting such analyses across multiple datasets, researchers can develop more robust evaluation methodologies and benchmarks that better foster the advancement of NL2VIS techniques.}

\revise{The limitation analysis of nvBench reveals key challenges in NL2VIS benchmarks, which may also exist in other datasets with varying degrees of impact.
Such analysis provides a foundation for systematically identifying issues across benchmarks and deepening understanding of NL2VIS evaluation weaknesses.
Extending this approach can support the creation of more reliable evaluation methods and benchmarks, ultimately advancing NL2VIS research.}
\revise{We also suggest that LLMs are promising as a tool for debugging and quality control of NL2VIS datasets.}

\section{Discussion and Future work}

\textbf{Further refinements and explorations for understanding LLM's capability.} While our work provides valuable insights into NL2VIS capabilities leveraging LLMs and steps forward in evaluating NL2VIS performance with LLMs, many opportunities exist for further refinements and explorations of understanding the LLMs' capability. 

\textit{(a) Consideration of alternative visualization construction methods:} While Vega-Lite serves as a popular choice for visualization construction, other methods like D3 offer distinct advantages and challenges. Future research should explore the applicability of NL2VIS approaches in a broader range of visualization paradigms, including a more flexible but challenging D3-based approach.

\delete{(b) Expansion beyond single grammar evaluation: Our evaluation focused solely on Vega-Lite grammar, overlooking the broader landscape of JSON-style visualization construction grammar. One existing study has surveyed over 57 JSON-style visualization specifications. Future investigations could extend to encompass diverse grammatical frameworks, to provide a more comprehensive understanding of NL2VIS capabilities.}

\textit{(b) Limitations of benchmark-based evaluation:} Prior analysis has noted that benchmark-driven methodologies may be too coarse-grained to surface deeper issues in generative visualization models~\cite{eurova24ws-Podo-framework4eval}. To address these limitations, future work could incorporate crowd-sourcing evaluations to provide complementary human-centered perspectives and richer insights into the effectiveness of NL2VIS systems.

\revise{\textbf{Futher directions for refining and broadening NL2VIS task:} To strengthen the reliability and coverage of NL2VIS task evaluation, several extensions can be pursued to address limitations. These directions aim to improve comparative fairness, mitigate data contamination concerns, and broaden the methodological foundation for future evaluation.}

\revise{\textit{(a) Establish a comparative framework for resources and generalizability:} We acknowledge that specialized, resource-efficient models, such as ncnet~\cite{Luo-tvcg2022-ncNet} achieves 60–80\% accuracy on nvBench, while requiring significantly lower computational and monetary costs than large proprietary LLMs used in ths study. 
However, these results are not directly comparable because ncnet is fine-tuned specifically on the nvBench dataset, whereas LLMs employ general-purpose understanding of LLM2VIS tasks using zero-shot and few-shot prompts. 
To enable a fair comparison, future research should establish a systematic evaluation framework that extends beyond simple accuracy. 
This framework should jointly assess the trade-offs in generalizability, accuracy, inference efficiency, and resource requirements. This approach will provide a balanced understanding of the true cost and benefit of LLMs versus smaller, specialized architectures.}

\revise{\textit{(b) Use fully open LLM models for deeper analysis.} While fully open LLM models such as TinyLlama could provide clearer guarantees regarding training-data transparency and contamination analysis, our current evaluation instead focuses on several widely adopted models of comparable scale to establish a practical and representative baseline for NL2VIS evaluation. Future work will extend the study to fully open-weight models, enabling more controlled contamination checks and a deeper interpretation of model behavior.}

\textbf{Future directions for improving NL2VIS performance based on LLMs.}
Our work can serve as a baseline for constructing data visualizations based on LLMs. In considering avenues for improving the performance of NL2VIS task using LLMs, several key areas emerge for further exploration.

\textit{(a) Specifying attribute sequences in generated Vega-Lite specifications:} Our experiments have revealed the influence of attribute sequence in the prompt on the resulting Vega-Lite specifications, as explained in Sec.~\ref{sec:evaluated-prompts}. One potential strategy for improvement involves incorporating additional rules governing attribute sequence generation. By refining the sequence specification process, we may enhance the coherence and quality of generated visualizations.

\textit{(b) Augmenting examples in generated specifications:} Prior research has demonstrated the efficacy of augmenting examples in enhancing model performance, particularly in tasks such as data wrangling. While our prompt design already incorporates a balanced selection of attributes when adding examples to the prompt, further investigation into the impact of increased example diversity is warranted. Future work could involve empirical validation of this approach to gauge its effectiveness in the NL2VIS tasks.

\textit{(c) Introducing multi-round user interactions:} The current evaluation focused on one-round generation to simulate end-to-end NL2VIS interactions. However, embracing a multi-round interaction paradigm holds promise for refining visualization construction. By enabling users to iteratively validate and refine attributes within generated specifications, we can potentially improve the accuracy and utility of LLM-generated visualizations.

\section{Conclusion}

\revise{In this work, we systematically evaluated six representative LLMs through eight prompt strategies on NL2VIS tasks using Vega-Lite specifications.
Our multi-metric assessment of vis accuracy, validity and legality reveals substantial performance differences across prompts, chart types and LLMs.
We also identified counterintuitive behaviors: more reasoning does not always lead to better performance, easier types do not always yield better results, and better-performing models do not always excel with every prompt.
Additionally, the nvBench exhibits benchmark limitations, including inaccurate query statements, inappropriate visual mapping and incorrect underlying data, introducing ambiguity into NL2VIS tasks.

These findings highlight the complexity of LLM interpretation in NL2VIS and underscore the need for improved prompting, more rigorous evaluation and cleaner benchmarks.
This study provides a foundation for developing more reliable, interpretable and practically deployable NL2VIS systems.
}

\bibliographystyle{IEEEtran}
\bibliography{main.bib}

\newpage

\newpage

\vfill

\end{document}